%% file: DBMC_NOMA_TMBMC_final_arxiv_package/paper.tex
\documentclass[lettersize,journal]{IEEEtran}
\usepackage{amsmath,amsfonts,amssymb}
\usepackage{algorithm}
\usepackage{array}
\usepackage{booktabs}
\usepackage[font=small,labelfont=bf]{caption}
\usepackage{subcaption}
\usepackage{textcomp}
\usepackage{stfloats}
\usepackage{url}
\usepackage{verbatim}
\usepackage{graphicx}
\usepackage{svg}
\usepackage[hidelinks]{hyperref} % load this BEFORE glossaries
\usepackage[acronym,shortcuts,nonumberlist]{glossaries} % nonumberlist = no page lists
\makeglossaries
\loadglsentries{abbreviations}  % your \newacronym definitions
\glsdisablehyper
\usepackage{cite}
\usepackage{lipsum}
\usepackage{siunitx}
\usepackage{algpseudocodex}
\usepackage{needspace}
\usepackage{titlesec}
\DeclareMathOperator*{\argmin}{arg\,min}

\DeclareMathAlphabet{\mathbbold}{U}{bbold}{m}{n}

\usepackage{supertabular}
\newglossarystyle{acr-2col-narrow}{%
  \renewcommand*{\glsgroupheading}[1]{}% no letter headings
}

\newif\ifrevision
\ifrevision
  \usepackage{xcolor}
  \newcommand{\rev}[1]{\textcolor{blue}{#1}} % new or modified text
  \newcommand{\revd}[1]{\textcolor{red}{\sout{#1}}} % deleted text (optional)
  \usepackage[normalem]{ulem} % for strikethrough
\else
  \newcommand{\rev}[1]{#1}
  \newcommand{\revd}[1]{}
\fi

\titlespacing\section{0pt}{5pt plus 2pt minus 2pt}{4pt plus 2pt minus 2pt}
\titlespacing\subsection{0pt}{3pt plus 1pt minus 1pt}{2pt plus 1pt minus 1pt}

\begin{document}

\title{DBMC-aNOMAly: Asynchronous NOMA with Pilot-Symbol Optimization Protocol for Diffusion-Based Molecular Communication Networks

\thanks{The authors acknowledge the financial support by the German Federal Ministry of Research, Technology and Space (BMFTR) in the program of “Souverän. Digital. Vernetzt.”. Joint project 6G-life, project identification number: 16KISK002.}
\thanks{Alexander Wietfeld and Wolfgang Kellerer are with the Chair of Communication Networks, Technical University of Munich, 80333 Munich, Germany (e-mail: \{alexander.wietfeld, wolfgang.kellerer\}@tum.de).}}% <-this % stops a space
%\thanks{Manuscript received April 19, 2021; revised August 16, 2021.}}

\author{Alexander Wietfeld,~\IEEEmembership{Graduate Student Member,~IEEE}, Wolfgang Kellerer,~\IEEEmembership{Fellow,~IEEE}
        % <-this % stops a space
}



\maketitle

\begin{abstract}
Multiple access (MA) schemes can enable cooperation between multiple nodes in future diffusion-based molecular communication (DBMC) networks. Non-orthogonal MA for DBMC networks (DBMC-NOMA) is a promising option for efficient simultaneous MA using a single molecule type. This paper studies parameter optimization and bit error probability (BEP) reduction for asynchronous DBMC-NOMA. First, we analytically derive the associated BEP and compare DBMC-NOMA with time-division and molecule-division MA. We show that asynchronous offsets can improve performance, and the upper-bound performance can be approached under almost all considered conditions by avoiding a small set of worst-case offset configurations, for which we propose and characterize a dedicated avoidance mechanism. We then propose DBMC-aNOMAly, a pilot-symbol-based optimization protocol for asynchronous DBMC-NOMA, and evaluate it using Monte Carlo simulations. DBMC-aNOMAly provides robust BEP reduction across different network sizes and noise levels, under sampling jitter, and under changing runtime conditions, outperforming protocols from previous work. An end-to-end efficiency analysis further shows that these gains translate into increased net throughput after compensating for the pilot overhead. DBMC-aNOMAly uses simple operations such as comparisons and additions that are compatible with chemical reaction networks, motivating future realistic modeling of the protocol. 
\end{abstract}

\begin{IEEEkeywords}
Molecular communication, asynchronous, non-orthogonal multiple access, optimization algorithm, protocol
\end{IEEEkeywords}

\section{Introduction}\label{sec:introduction}

\IEEEPARstart{M}{olecular} communication (MC) is a novel paradigm based on the transfer of information using molecules. Specifically, \ac{DBMC} is envisioned to play a major role as an energy-efficient and biocompatible approach at micro- or nano-scales~\cite{farsad_comprehensive_2016}.

A main vision for \ac{DBMC} is the \ac{IoBNT}~\cite{akyildizInternetBioNanoThings2015}, an interconnected network of tiny \acp{BNM} that extends the \ac{IoT} to the nano-scale using biological mechanisms such as \ac{DBMC}.
Predicted use cases range from the agricultural sector and laboratory settings to future medical approaches. For example, ongoing research is exploring the ideas of infection and tumor detection~\cite{gomezMachineLearningApproach2021}, cardiovascular system monitoring~\cite{hofmannMolecularCommunicationPerspective2024a}, or targeted drug delivery~\cite{chude-okonkwoMolecularCommunicationNanonetwork2017}.
It is widely understood that \acp{BNM} will largely be synthetic cells or tiny nano-robots with severely limited resources and capabilities~\cite{maitraInternetHarvesterNano2025, kocaInformationTheoreticLifetimeMaximization2025}.
On the other hand, a large number of \acp{BNM} will be deployed at once. Therefore, realizing these use cases requires \ac{DBMC} networks for communication and collaboration between \acp{BNM}~\cite{wietfeldDBMCNOMAEvaluatingNOMA2024, bienauMolecularCommunication6G2025}.
Additionally, systems within the \ac{IoBNT} lack the general computing capabilities that are the foundation of traditional \ac{EM}-based networks. These systems require adapted and simplified computing approaches so that individual \acp{BNM} can follow algorithm steps, make decisions and optimize parameters~\cite{heinleinClosingImplementationGap2024, angerbauerMolecularNanoNeural2024, wietfeldErrorProbabilityOptimization2024c, liuDNADecisionTree2025}. Studies in physiological MC channels emphasize that environmental effects can distort signal features, motivating simple signals and processing~\cite{felicetti2025BloodVessels}.

An overview of all acronyms used throughout the paper can be found in Table~\ref{tab:acronyms}.

\begin{table}[hb]
\vspace{-0.0cm}
\caption{List of Acronyms}
\vspace{-0.1cm}
\label{tab:acronyms}
\centering
\footnotesize
\setlength{\tabcolsep}{3pt}
\renewcommand{\arraystretch}{0.96}
\begin{tabular}{@{}>{\raggedright\arraybackslash}p{0.28\columnwidth} >{\raggedright\arraybackslash}p{0.68\columnwidth}@{}}
\toprule
\textbf{Acronym} & \textbf{Full term} \\
\midrule
BEP & bit error probability \\
BNM & Bio-Nano-Machine \\
CRN & chemical reaction network \\
DBMC & diffusion-based molecular communication \\
DBMC-aNOMAly & \parbox[t]{0.68\columnwidth}{asynchronous NOMA with pilot-symbol optimization\\protocol for DBMC} \\
DBMC-NOMA & NOMA for DBMC networks \\
EM & electromagnetic \\
IoBNT & Internet of Bio-Nano-Things \\
IoT & Internet of Things \\
ISI & inter-symbol interference \\
MA & multiple access \\
MAI & multiple-access interference \\
MDMA & molecule-division multiple access \\
MI & mutual information \\
NOMA & non-orthogonal multiple access \\
OOK & on-off-keying \\
RX & receiver \\
SIC & successive interference cancellation \\
SNR & signaling-molecule-to-noise ratio \\
TDMA & time-division multiple access \\
TX & transmitter \\
UCA & uniform concentration assumption \\
WCAM & worst-case-offset avoidance mechanism \\
\bottomrule
\end{tabular}
\vspace{-0.0cm}
\end{table}

% \needspace{2\baselineskip}

\begin{figure}[t]
    \centering
    \includegraphics[width=0.95\linewidth]{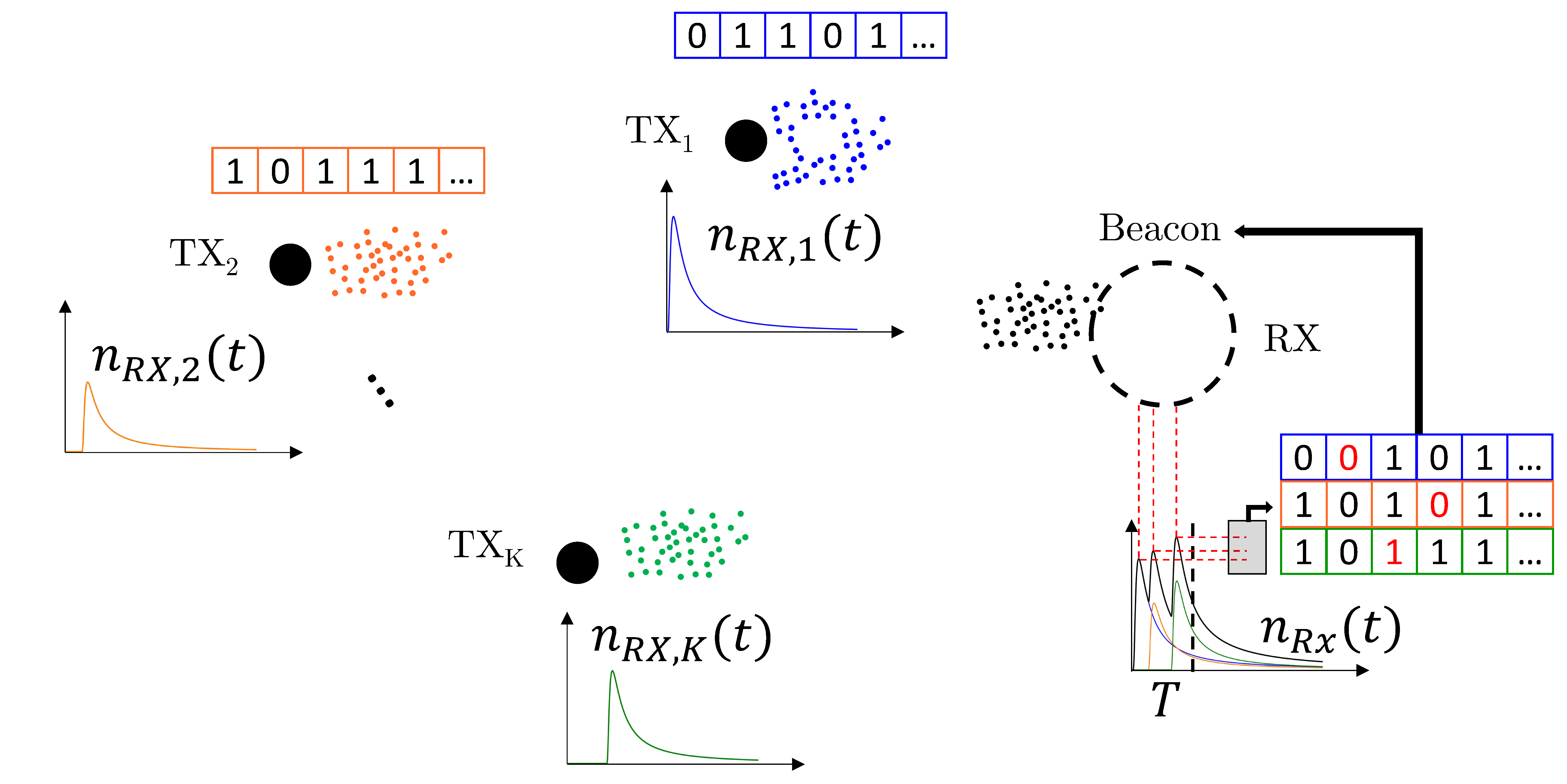}
    \caption{Proposed asynchronous DBMC-NOMA protocol (DBMC-aNOMAly), including \acs{NOMA} transmission from many \acsp{TX} to one \acs{RX}, pilot-symbol-based parameter optimization, and the \acs{RX} beacon for avoiding worst-case offset scenarios.}
    \label{fig:overview}
    \vspace{-0.3cm}
\end{figure}

The first step towards a comprehensive \ac{DBMC} networked system is an efficient \ac{MA} scheme. \Ac{MA} can ensure that a \ac{RX} \ac{BNM} can differentiate the content and sources of messages arriving from multiple \ac{TX} \acp{BNM}.
In this paper, we extend our previous \ac{DBMC-NOMA} baseline~\cite{wietfeldDBMCNOMAEvaluatingNOMA2024} to the practically relevant asynchronous setting and analyze its performance for multiple \acp{TX}. Building on these insights, we propose DBMC-aNOMAly, a simple pilot-symbol-based protocol for parameter optimization and robust operation under random and time-varying synchronization offsets between \acp{TX}. A simple schematic representation is shown in Fig.~\ref{fig:overview}.

Potential applications include many-to-one \emph{data-gathering} and \emph{status-update} scenarios, where a large number of low-capability \acp{BNM} report short messages to a single gateway \ac{RX} \ac{BNM}. Two representative examples are: (i) \emph{in-body collaborative sensing}, where swarms of nanosensors monitor biomarkers and jointly report to an \ac{RX} for detection and actuation; and (ii) \emph{microfluidic and lab-on-chip platforms}, where multiple reaction compartments periodically report outcomes to a shared readout port~\cite{hamidovicMicrofluidicSystemsMolecular2024, pappalardoSyntheticMolecularCommunication2025}. Allocating a dedicated molecule type per \ac{TX} (as in \ac{MDMA}) scales poorly with receptor complexity and biochemical compatibility, while tight global synchronization (as in \ac{TDMA}) is challenging~\cite{krishnaswamyADMAAmplitudeDivisionMultiple2017a}. Allowing multiple \acp{TX} to transmit within the same time interval increases the number of reports collected per unit time and improves the achievable update rate at the gateway. 
This matches the classical motivation of power-domain \ac{NOMA}, where multiple users share the same resource and are separated at the receiver to improve bandwidth efficiency and support massive connectivity~\cite{saitoNOMA2013,daiSurveyNonOrthogonalMultiple2018}. These constraints motivate a single-molecule-type \ac{MA} scheme that supports concurrent transmissions and tolerates asynchronous offsets, which is precisely the operating regime targeted by asynchronous \ac{DBMC-NOMA}~\cite{wietfeldDBMCNOMAEvaluatingNOMA2024, wietfeldErrorProbabilityOptimization2024c} and the proposed DBMC-aNOMAly protocol.
% \vspace{-0.5\baselineskip}
\subsection{Related Work}
Here, we review \ac{MA} schemes for \ac{DBMC} and prior parameter-optimization methods, then identify the remaining gaps and our contributions.
This paper is an extension of our previous work on \ac{DBMC-NOMA}~\cite{wietfeldDBMCNOMAEvaluatingNOMA2024} and error probability optimization for \ac{MA}~\cite{wietfeldErrorProbabilityOptimization2024c}.
\subsubsection{Multiple Access for DBMC}
Several \ac{MA} schemes have been proposed and analyzed for \ac{DBMC} networks. Each is based on utilizing a physical resource to be distributed to users. We will view the schemes from the perspective of a network with a single \ac{RX} and multiple \acp{TX}.

\Ac{TDMA} utilizes the time dimension, sequentially assigning a time slot to each \ac{TX}, in which it can transmit a message~\cite{shitiriTDMABasedDataGathering2021, rudsariDrugReleaseManagement2019}, requiring all \acp{TX} to be tightly synchronized with each other but allowing the same molecule type for transmission.

The \ac{MDMA} scheme assigns a different molecule type to each \ac{TX}, making it possible for the \ac{RX} to differentiate the messages into separate streams, even if they arrive simultaneously~\cite{chouhanRescaledBrownianMotion2022, chenResourceAllocationMultiuser2021}, enabling parallel signaling from all \acp{TX}. However, growing numbers of different molecules are necessary for larger networks and the physical complexity of the \ac{RX} grows, as more receptor types are required, either linearly, or sub-linearly for mixture-based extensions~\cite{jamaliOlfactioninspiredMCsMolecule2023}.

The use of the amplitude dimension for \ac{MA} in \ac{DBMC} was first proposed in~\cite{krishnaswamyADMAAmplitudeDivisionMultiple2017a}, where the number of received molecules from each \ac{TX} is defined as a source address and a set of amplitudes is designed such that the sum of a certain subset is unique to a set of active \acp{TX}. However, the system lacks the capability to adapt to random or changing conditions and the assumptions of the underlying physical scenario are not tailored to real \ac{DBMC} networks.

\Ac{NOMA} has received significant attention in the context of classical \ac{EM} communication in recent years~\cite{saitoNOMA2013}, due to its capability to increase the network capacity without requiring more frequency channels. Power-domain \ac{NOMA}, its most common form, is based on different levels of received signal power at the \ac{RX}. If we compare \ac{EM} communication and \ac{DBMC}, there are analogies between frequency channels and molecule types, as well as signal power and the received number of molecules. This motivates transferring the \ac{NOMA} principle to \ac{DBMC} to enable concurrent transmissions with a single molecule type and receiver-side separation (e.g., via \ac{SIC}), avoiding the molecule-type scaling of \ac{MDMA}.

The \ac{DBMC-NOMA} scheme from~\cite{wietfeldDBMCNOMAEvaluatingNOMA2024} has been shown to match the performance of \ac{MDMA}, as long as communication parameters including detection thresholds and the emitted number of molecules are chosen optimally. However, the previous work only considered the case where all \acp{TX} are synchronized with each other, simplifying the mathematical analysis, but leading to severe performance limitations. 
% Similar to discussions for traditional \ac{NOMA}~\cite{ganjiTimeAsynchronousNOMA2019}, and as we will show in this paper, extending the analytical model to the asynchronous case between all \acp{TX} can yield significant performance improvements by reducing the \ac{MAI}.
Additionally, the work in~\cite{wietfeldDBMCNOMAEvaluatingNOMA2024} did not address the question of implementation hurdles for \ac{NOMA}, particularly, the optimal choice of parameters in a practical system and the increased complexity due to the \ac{SIC} mechanism.

\subsubsection{Parameter Optimization for DBMC Networks}

Several approaches optimize \ac{DBMC} parameters to minimize \ac{BEP} or related metrics. Exhaustive search over the analytical expression can find the global optimum with sufficient resolution and an appropriate search domain~\cite{wietfeldDBMCNOMAEvaluatingNOMA2024}, but becomes infeasible in practical systems or larger networks. Alternatives include closed-form solutions~\cite{chouhanOptimalTransmittedMolecules2019}, global optimization such as gradient descent and particle swarm optimization~\cite{chengOptimizationDecisionThresholds2022}, and data-driven threshold selection~\cite{qianMolecularCommunicationsModelBased2019a}. However, these methods require tractable or repeatedly evaluated \ac{BEP} expressions, accurate model inputs, or implementation overhead that is difficult to reconcile with limited \ac{TX}/\ac{RX} resources and biochemical realizability constraints~\cite{angerbauerMolecularNanoNeural2024}.

\subsubsection{Chemically-Compatible Heuristic Methods}

Lastly, specific heuristic algorithms have been proposed. Pilot-based adaptation using simple primitives (comparisons, additions, etc.) is attractive because such operations can be realized with CRNs~\cite{vasicCRNMolecularProgramming2020}, and fully chemical synchronization/detection has been demonstrated~\cite{heinleinClosingImplementationGap2024}. 
Recent work on DNA-strand-displacement has shown that multi-level decision trees can be implemented chemically with high precision for up to 10 layers or more~\cite{liuDNADecisionTree2025}, similar to the \ac{SIC} decision tree implementation we will present in Section~\ref{sec:system_model}. In our previous work, we have proposed a first version of an optimization heuristic for the \ac{DBMC-NOMA} scheme~\cite{wietfeldErrorProbabilityOptimization2024c} and have simulated a version based on \acp{CRN}~\cite{wietfeldChemSICalEvaluatingStochastic2025}. However, the system in~\cite{wietfeldErrorProbabilityOptimization2024c} is limited to only 2 \acp{TX} and explores only a few variable parameters. Additionally, the system model in~\cite{wietfeldErrorProbabilityOptimization2024c} assumes perfect global synchronization, i.e., no offset between the peak molecule arrival times from all \acp{TX}, and relies on a complex feedback channel to optimize the emitted molecule counts. This paper addresses both limitations.

\subsection{Contributions and extension of previous work}

In this work, we build upon our previous work~\cite{wietfeldDBMCNOMAEvaluatingNOMA2024, wietfeldErrorProbabilityOptimization2024c} to make the following novel contributions:

\begin{enumerate}
    \item We present a comprehensive analytical model for an asynchronous \ac{NOMA} system for \ac{DBMC} networks with $K$ \acp{TX}, one \ac{RX}, incorporating $L$ symbols of \ac{ISI}.
    \item We derive the analytical \ac{BEP} expression of this system assuming a diffusion-based Poisson channel.
    \item Using the analytical results, we present a comparison between \ac{TDMA}, \ac{MDMA} and \ac{DBMC-NOMA} focusing specifically on different scenarios of synchronization offset. We show that effectively utilizing the offset can significantly improve the performance of \ac{DBMC-NOMA}, outperforming \ac{TDMA} and matching \ac{MDMA} in much harsher conditions compared to the synchronized \ac{DBMC-NOMA} scheme considered in~\cite{wietfeldDBMCNOMAEvaluatingNOMA2024} and~\cite{wietfeldErrorProbabilityOptimization2024c}, as long as certain worst-case scenarios are avoided.
    \item Furthermore, we propose a pilot-symbol-based optimization protocol for asynchronous \ac{DBMC-NOMA} (DBMC-aNOMAly), capable of adaptively choosing detection thresholds, optionally adapting the number of transmitted molecules, and including a \ac{WCAM}. By explicitly accounting for the individual synchronization offsets, the protocol represents a significant extension of the version considered in~\cite{wietfeldErrorProbabilityOptimization2024c} and is based on simple operations compatible with \ac{CRN} implementation.
    \item Lastly, we analyze DBMC-aNOMAly in terms of convergence and practical communication efficiency under different conditions. We show robust \ac{BEP} reduction for different noise levels, sampling jitter, changing channel conditions, and larger networks with $K>2$. We also characterize the \ac{WCAM} by trigger behavior, offset selectivity, and immediate post-trigger \ac{BEP} reduction, and show that the protocol yields net throughput gains after accounting for pilot overhead while making emitted-molecule-count optimization unnecessary in most cases.
\end{enumerate}

\noindent Our previous work~\cite{wietfeldDBMCNOMAEvaluatingNOMA2024} analyzed a \emph{synchronized} \ac{DBMC-NOMA} baseline, and~\cite{wietfeldErrorProbabilityOptimization2024c} proposed a pilot-based heuristic for \emph{$K=2$ under global synchronization} with a molecule-count feedback loop. The present paper (i) develops an \emph{asynchronous} \ac{DBMC-NOMA} model with per-TX offsets and $K$ streams, (ii) characterizes \emph{worst-case offset regions} that explain when \ac{NOMA} falls short and how \ac{MDMA}-like performance is recovered, and (iii) proposes DBMC-aNOMAly, which optimizes the \ac{BEP} via pilot-based threshold adaptation and \ac{WCAM} without explicit channel estimation or tight network-wide synchronization.

The remainder of this paper is structured as follows. Section~\ref{sec:system_model} introduces the channel model, communication assumptions, and asynchronous models for \ac{TDMA}, \ac{MDMA}, and \ac{DBMC-NOMA}. Section~\ref{sec:bep} derives the analytical \ac{BEP}, and Section~\ref{sec:ma_comparison} evaluates the \ac{MA} schemes. Section~\ref{sec:protocol} presents DBMC-aNOMAly, Section~\ref{sec:protocol_eval} evaluates the protocol by simulation, and Section~\ref{sec:conclusion} concludes the paper.

\section{Scenario and System Model}\label{sec:system_model}

This section introduces the physical scenario, defines the channel model, and describes the communication system and considered \ac{MA} schemes.

\begin{figure}[t]
    \centering
    \includegraphics[width=0.85\linewidth]{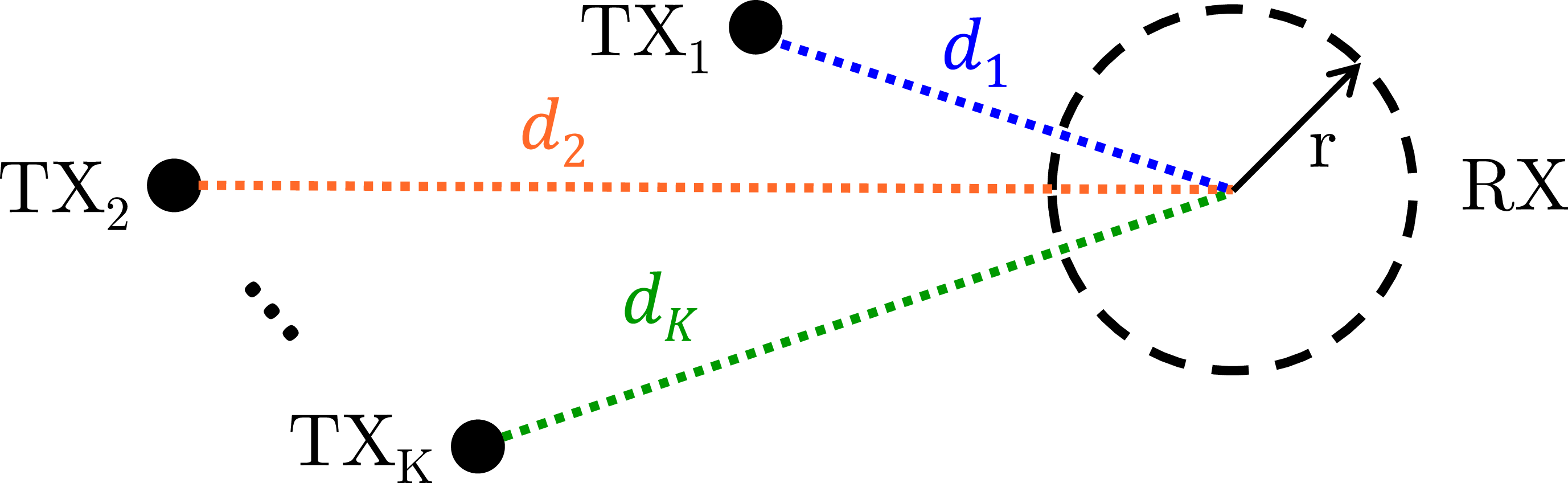}
    \caption{DBMC scenario with $K$ point-source \acp{TX} at distances $d_1$, $d_2$, $\dots$, $d_K$ from a spherical \ac{RX}.}
    \label{fig:scenario}
\end{figure}

\subsection{Channel Model}

 Fig.~\ref{fig:scenario} depicts the considered basic communication scenario. The network consists of a single \ac{RX} and $K$ \acp{TX} at distances $d_1, d_2, \dots, d_K$, where TX$_i$ is at distance $d_i$ from the \ac{RX}.
The \ac{RX} is assumed to be a passive spherical observer with radius $r$, representing a common simple \ac{RX} model~\cite{jamaliChannelModelingDiffusive2019}.
The \acp{TX} are modeled as point sources which are capable of emitting instantaneous pulses of molecules. This represents a good approximation compared to a volume \ac{TX} as long as the channel is sufficiently long~\cite{jamaliChannelModelingDiffusive2019}.
The surrounding channel is assumed to be unbounded free space. Accordingly, free diffusion affects the emitted molecules with diffusion coefficient $D$.
Firstly, we will focus on the single-link channel between a TX$_i$ and the \ac{RX}. If TX$_i$ releases a single molecule at $t=0$, we can solve the diffusion equation to find the probability of observing this molecule within the \ac{RX} volume $V_\mathrm{RX} = \frac{4}{3}\pi r^3$ at time $t$ as~\cite{jamaliChannelModelingDiffusive2019}
    \begin{equation}\label{eq:P_t}
        P_i(t) = P(t,d_i) = \frac{V_\mathrm{RX}}{\left(4\pi Dt\right)^\frac{3}{2}}\exp \left(-\frac{d_i^2}{4Dt}\right).
    \end{equation}
For (\ref{eq:P_t}), we apply the \ac{UCA} inside the \ac{RX}, which is approximately valid for sufficiently long channels compared to the \ac{RX} size, i.e. $r<0.15\cdot d_i$~\cite{jamaliChannelModelingDiffusive2019}.
Now, we consider that TX$_i$ releases $N_{\mathrm{TX},i}$ molecules simultaneously and $N_{\mathrm{TX},i}$ is sufficiently large compared to the received number of molecules $n_{\mathrm{RX},i}(t)$. Additionally, as is typical~\cite{jamaliChannelModelingDiffusive2019}, the propagation of individual molecules is assumed independent. Then, $n_{\mathrm{RX},i}(t)$ can be modeled as a Poisson distributed random variable $n_{\mathrm{RX},i}(t)\sim \mathcal{P}(\lambda_i(t))$ with mean and variance equal to 
    \begin{equation}\label{eq:lambda}
        \lambda_i(t) = N_{\mathrm{TX},i}P_i(t).
    \end{equation}

\subsection{Communication System}\label{subsec:communication_system}

For the modulation scheme, we utilize \ac{OOK}, i.e. $N_{\mathrm{TX},i}$ molecules are released for a bit-1 and nothing for a bit-0.
Time is split into slots of length $T$. The current time slot, starting at $t=0$ and ending at $t=T$ is denoted as slot $l=0$, while the preceding $L$ slots are denoted as $l\in \{1,2,\dots, L\}$.
The symbol sent by each TX$_i$ in time slot $l$ is $s_i[l]\in\{0,1\}$ with 0 and 1 equally likely.
Every \ac{TX} emits a pulse of $s_i[l]N_{\mathrm{TX},i}$ molecules at time $t_{\mathrm{off},i}$ within the current time slot, where $0\leq t_{\mathrm{off},i}<T$.
Therefore, as opposed to previous work in~\cite{wietfeldDBMCNOMAEvaluatingNOMA2024, wietfeldErrorProbabilityOptimization2024c}, we do not generally assume synchronization between the \acp{TX}, which could be excessively complicated, particularly as the number of \acp{TX} $K$ increases. Also, in contrast to~\cite{wietfeldDBMCNOMAEvaluatingNOMA2024}, we do not need accurate channel knowledge and distance information at the \ac{RX}.

However, we require that the \ac{RX} acquires an estimated sampling time for each \ac{TX}. This could be possible in practice by having multiple oscillating \acp{CRN} within the \ac{RX} that react to a certain pulse trigger sent by each \ac{TX} or from an external source. A similar mechanism is described in~\cite{heinleinClosingImplementationGap2024, borgesSynchronizationProtocolMultiUser2021, jiangClockSignalGeneration2025}. We will not model this process explicitly. Instead, we capture its residual timing uncertainty by a bounded sampling-time jitter model. We denote all \ac{RX}-side processes associated with a certain \ac{TX} by the index $j$ to differentiate it from the sending side. Let $t_{\mathrm{p},j} = \frac{d_j^2}{6D} + t_{\mathrm{off},j}$ denote the channel-determined peak time of the expected response for TX$_j$. The \ac{RX} is \emph{not} assumed to know $d_j$ or $t_{\mathrm{p},j}$, but only to obtain an estimate $t_{\mathrm{s},j}$. Following common practice in \ac{DBMC} synchronization studies that model unknown timing within a known acquisition window as uniform~\cite{jamaliSymbolSynchronizationDiffusive2017}, we assume
\begin{equation}
    t_{\mathrm{s},j}\sim \mathcal{U}(t_{\mathrm{p},j}-\Delta_\mathrm{p}/2,\, t_{\mathrm{p},j}+\Delta_\mathrm{p}/2),
\end{equation}
with sampling jitter $\Delta_\mathrm{p}$.

Throughout the paper, TX$_j$ denotes a \emph{logical stream} at the \ac{RX} that is associated with one physical \ac{TX} via the sampling-time acquisition mechanism. Concretely, we assume an initialization phase in which each \ac{TX} emits a short trigger (e.g., on a dedicated control molecule type or using a time-coded preamble) that allows the \ac{RX} to (i) instantiate and maintain a \ac{TX}-specific internal timing process and (ii) assign a persistent index $j\in\{1,\dots,K\}$ to that \ac{TX}. Hence, the \ac{RX} can associate samples taken at $t_{\mathrm{s},j}$ with TX$_j$ over time. We may not require semantic source identities due to commonly assumed higher-layer data-gathering applications in the \ac{IoBNT}~\cite{shitiriTDMABasedDataGathering2021}, where the exact origin of each message is not needed.

The average signal component sent by TX$_i$ for a bit-1 and sampled at the sampling time of TX$_j$ in time slot $l$ is
    \begin{equation}
        \lambda_{i,j}[l] = \lambda_i(t_{\mathrm{s},j} + lT).
    \end{equation}
Additionally, we denote the \textit{desired average sample}, i.e. the average contribution at $t_{\mathrm{s},i}$ for a bit-1 from TX$_i$ in the current time slot as
    \begin{equation}
        \tilde{\lambda}_i = \lambda_{i,i}[0] = \lambda_i(t_{\mathrm{s},i}).
    \end{equation}
The total signal at the \ac{RX} for a certain sampling point, $n_\mathrm{RX}(t_{\mathrm{s},j})$, is a sum of multiple independent Poisson random variables $n_{\mathrm{RX},i}(t_{\mathrm{s},j})$ with means either equal to 0 or according to (\ref{eq:lambda}). Therefore, the sum is also a Poisson random variable with its mean the sum of the means of the added random variables. Unintended leakage from molecule reservoirs in the \ac{TX} or \ac{RX}, and interference from other unrelated surrounding processes represent an additional source of molecules at the \ac{RX}~\cite{noelUnifyingModelExternal2014}. Consistent with models for continuous noise sources in \ac{DBMC} channels, we assume an additive Poisson noise component with mean $\lambda_\mathrm{n}$ to represent these environmental noise factors~\cite{noelUnifyingModelExternal2014}. The calculation of the entire sum depends on the offsets $t_{\mathrm{off},i}$ and the utilized \ac{MA} scheme and will be detailed in the following section.

\subsection{Multiple Access Schemes}

\begin{figure}[t]
    \centering
    \includegraphics[width=\linewidth]{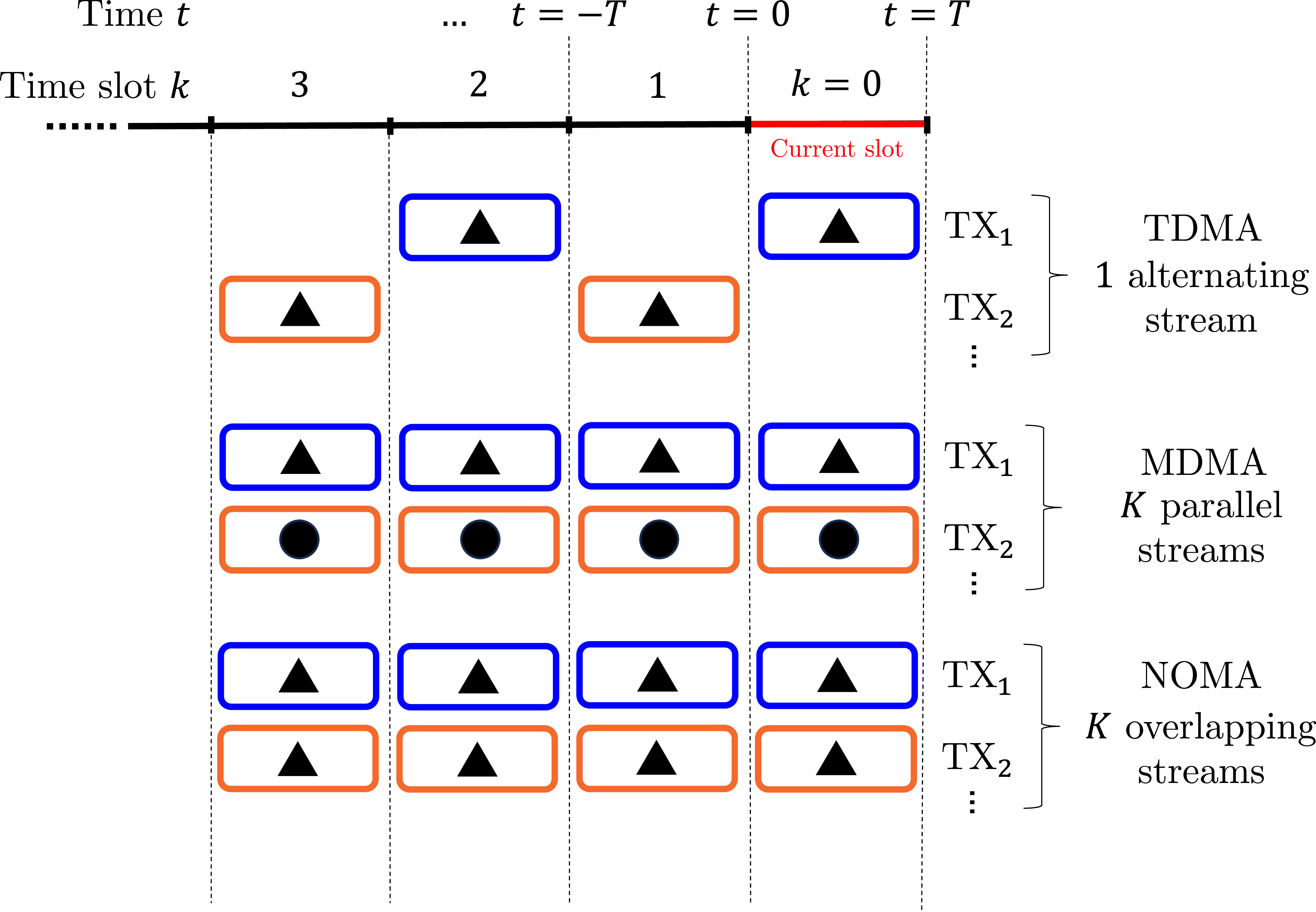}
    \caption{Comparison of resource allocation for different MA schemes. The triangle and circle shape signify different molecule types, the color is associated with different \acp{TX}, and the horizontal placement in the time slots shows the allocation of transmissions over time.}
    \label{fig:ma_schemes}
\end{figure}

We will consider \ac{MDMA}, \ac{TDMA}, and \ac{DBMC-NOMA} for the analytical evaluation. The calculation of the average received signal for the Poisson distribution will be derived for each scheme. An overview of the resource allocation using time and molecule type is shown in Fig.~\ref{fig:ma_schemes} for illustrative purposes.

\paragraph{Molecule-Division Multiple Access (MDMA)}

Starting with \ac{MDMA}, each \ac{TX} utilizes one of $K$ different molecule types. We assume that they are perfectly distinguishable at the \ac{RX} such that $K$ parallel channels are created, which we can analyze separately. 
It is assumed that each \ac{TX} transmits a symbol in every time slot $l$ and that all molecule types share the diffusion coefficient $D$.
We can consider the signal received from any TX$_j$ without loss of generality, and it can be expressed as
    \begin{equation}\label{eq:n_mdma}
        n_{\mathrm{RX},j}^\mathrm{MDMA}(t_{\mathrm{s},j}) \sim \mathcal{P}\biggl(\lambda_\mathrm{n} + s_j[0]\tilde{\lambda}_j + \underbrace{\sum_{l=1}^{L}s_j[l]\lambda_{j,j}[l]}_\mathrm{ISI} \biggr)
    \end{equation}
We can see that this contains the \textit{desired sample} component $s_j[0]\tilde{\lambda}_j$ as well as \ac{ISI}.
Then, the sample taken by the \ac{RX} to detect the symbol sent by TX$_j$ in the current time slot is $n_{\mathrm{s},j}^\mathrm{MDMA} =  n_{\mathrm{RX},j}^\mathrm{MDMA}(t_{\mathrm{s},j})$.
Threshold detection is applied according to
    \begin{equation}\label{eq:detection_mdma}
        \hat{s}_j=
        \begin{cases}
            1 & n_{\mathrm{s},j}^\mathrm{MDMA} \geq \tau_j^\mathrm{MDMA}\\
            0 & n_{\mathrm{s},j}^\mathrm{MDMA} < \tau_j^\mathrm{MDMA},
        \end{cases}
    \end{equation}
with the detected symbol $\hat{s}_j$ and the set of thresholds $\mathcal{T}^\mathrm{MDMA} = \{\tau_j^\mathrm{MDMA}\}_{j=1}^{K}$.

\paragraph{Time-Division Multiple Access (TDMA)}

In \ac{TDMA}, we assume all \acp{TX} use the same molecule type and they are assigned different time slots in sequential order.
When the current time slot is assigned to TX$_j$, then each previous time slot $l>0$ was assigned to \ac{TX}$_{[(j-l-1)\ \mathrm{mod}\ K] + 1}$, i.e. continuing in descending order and looping around to TX$_K$ at $l=j$.
 Then, the received signal is the sum of all signals emitted by the \acp{TX} and, sampled at $t_{\mathrm{s},j}$, is given by
    \begin{multline}\label{eq:n_tdma}
        n_{\mathrm{RX}}^\mathrm{TDMA}(t_{\mathrm{s},j}) \sim 
            \mathcal{P}\biggl(\lambda_\mathrm{n} + s_j[0]\tilde{\lambda}_j \cdots \\ 
            \cdots + \underbrace{\sum_{l=1}^{L}s_{[(j-l-1)\ \mathrm{mod}\ K] + 1}[l]\lambda_{[(j-l-1)\ \mathrm{mod}\ K] + 1,j}[l]}_\mathrm{ISI} \biggr)
    \end{multline}
It includes \ac{ISI} from previous time slots, in which various other \acp{TX} were transmitting.
To detect the symbol transmitted by TX$_j$, assigned to the current time slot, we use the sample $n_{\mathrm{s},j}^\mathrm{TDMA} = n_{\mathrm{RX}}^\mathrm{TDMA}(t_{\mathrm{s},j})$ and apply detection
    \begin{equation}\label{eq:detection_tdma}
        \hat{s}_j=
        \begin{cases}
            1 & n_{\mathrm{s},j}^\mathrm{TDMA} \geq \tau_j^\mathrm{TDMA}\\
            0 & n_{\mathrm{s},j}^\mathrm{TDMA} < \tau_j^\mathrm{TDMA},
        \end{cases}
    \end{equation}
with the set of detection thresholds $\mathcal{T}^\mathrm{TDMA} = \{\tau_j^\mathrm{TDMA}\}_{j=1}^{K}$.

\paragraph{Non-Orthogonal Multiple Access for DBMC Networks (DBMC-NOMA)}

We utilize and extend the initial definition of \ac{DBMC-NOMA} from our previous work~\cite{wietfeldDBMCNOMAEvaluatingNOMA2024, wietfeldErrorProbabilityOptimization2024c}.
All \acp{TX} transmit in each time slot using the same molecule type, as shown in Fig.~\ref{fig:ma_schemes}.
The received signal at the sampling time $t_{\mathrm{s},j}$ for \ac{TX}$_j$ will include contributions from past transmissions, i.e. \ac{ISI}, as well as from other \acp{TX} in the same time slot, i.e. \ac{MAI}.
 The resulting formula is given by
    \begin{multline}\label{eq:n_rx_noma}
        n_\mathrm{RX}^\mathrm{NOMA}(t_{\mathrm{s},j}) \\ \sim \mathcal{P}\biggl(\lambda_\mathrm{n} + s_j[0]\tilde{\lambda}_j + \underbrace{\sum_{\substack{i=1 \\ i \neq j}}^{K} s_i[0]\lambda_{i,j}[0]}_{\mathrm{MAI}} + \underbrace{\sum_{i=1}^{K}\sum_{l=1}^{L}s_i[l]\lambda_{i,j}[l]}_{\mathrm{ISI}}\biggr).
    \end{multline}

To reduce the impact of \ac{MAI}, the \ac{RX} applies \ac{SIC} and decodes the $K$ streams sequentially from TX$_1$ to TX$_K$ within each time slot. For stream $j$, the \ac{RX} uses the sample
\begin{equation}
    n_{\mathrm{sample},j}^\mathrm{NOMA} = n_\mathrm{RX}^\mathrm{NOMA}(t_{\mathrm{s},j}),
\end{equation}
and makes a binary decision.

\emph{Classical mean-cancellation view:}
In conventional power-domain \ac{NOMA}, after deciding $\hat{s}_i$ for users $i<j$, the \ac{RX} cancels their \emph{expected} contributions from the observation before decoding user $j$. In our \ac{DBMC} setting, this corresponds to forming a residual sample
\begin{equation}
    \tilde{n}_j = n_{\mathrm{sample},j}^\mathrm{NOMA} - \sum_{i=1}^{j-1} \hat{s}_i \lambda_{i,j}[0],
\end{equation}
and comparing $\tilde{n}_j$ to a baseline threshold.

\begin{figure}[t]
    \centering
    \includegraphics[width=\linewidth]{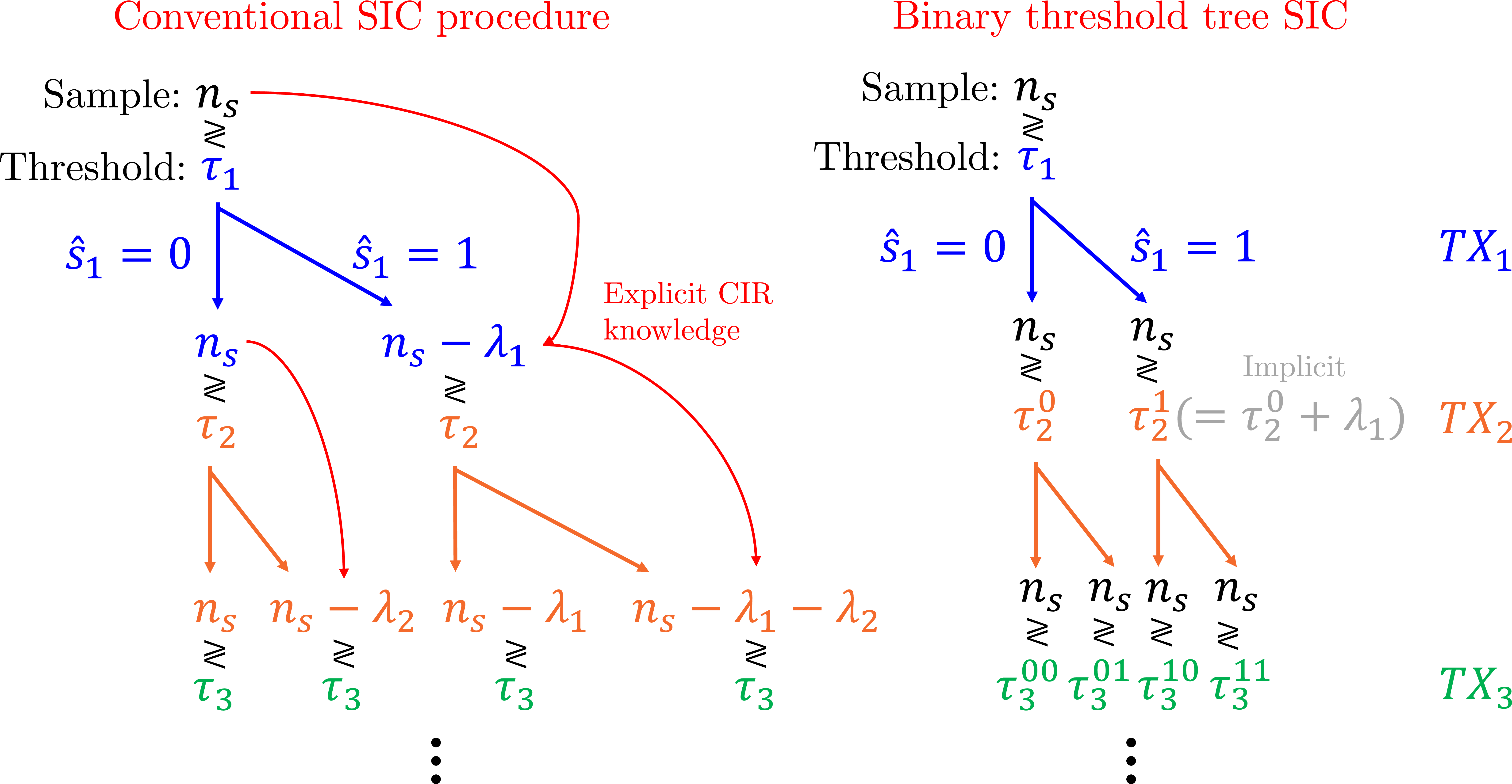}
    % \missingfigure[figwidth=0.8\linewidth]{SIC method comparison between additive threshold determination using channel knowledge and binary tree multi-threshold detection}
    \caption{\Acf{SIC} implementation strategies: the diagram on the left shows the classical description of a \ac{SIC} procedure, using knowledge of the individual channel impulse response to subtract the respective contribution from TX$_i$, $\lambda_i$, from the sampled signal $n_\mathrm{s}$. On the right, the equivalent threshold-adaptation-based version is shown, which uses a set of thresholds $\tau_i^{\hat{s}_{i-1}}$ for every \ac{TX} resulting in a binary-tree-like decision structure.}
    \label{fig:sic_comparison}
\end{figure}

\emph{Equivalent threshold-adaptation view (binary-tree detection):}
Equivalently, the same decision can be written \emph{without explicitly forming} $\tilde{n}_j$ by shifting the threshold depending on the previously detected vector $\mathbf{\hat{s}}_{j-1} = [\hat{s}_1,\dots,\hat{s}_{j-1}]$. Thus, for TX$_j$ we use
\begin{equation}\label{eq:detection_noma}
    \hat{s}_j = \begin{cases}
    1 & n_{\mathrm{sample},j}^\mathrm{NOMA} \geq \tau_j^{\mathbf{\hat{s}}_{j-1}} \\
    0 & n_{\mathrm{sample},j}^\mathrm{NOMA} < \tau_j^{\mathbf{\hat{s}}_{j-1}}.
    \end{cases}
\end{equation}
Here, the set $\mathcal{T}_j^\mathrm{NOMA} = \{\tau_j^{0\dots 00}, \tau_j^{0\dots 01}, \dots, \tau_j^{1\dots 11}\}$ contains $2^{j-1}$ possible thresholds, which form a binary decision tree as illustrated in Fig.~\ref{fig:sic_comparison}. Importantly, during runtime only \emph{one} branch is traversed, namely the branch indexed by the obtained $\mathbf{\hat{s}}_{j-1}$. This formulation is algebraically equivalent to mean-cancellation when $\tau_j^{\mathbf{\hat{s}}_{j-1}}$ is chosen as a baseline threshold plus the corresponding sum of canceled mean contributions, but it also allows treating $\tau_j^{\mathbf{\hat{s}}_{j-1}}$ as a free parameter for pilot-based optimization, avoiding explicit channel estimation.

Since the selected threshold depends on previous decisions, \ac{SIC} is inherently affected by error propagation. This effect is explicitly accounted for in our \ac{BEP} derivation by averaging over all possible $\mathbf{\hat{s}}_{j-1}$ branches in Section~\ref{sec:bep}.

\emph{Scaling and feasibility:} In the analytical representation of Eq.~(\ref{eq:detection_noma}), TX$_j$ has $2^{j-1}$ possible decision-conditioned thresholds, such that the full detector contains $\sum_{j=1}^{K} 2^{j-1} = 2^K-1$ threshold contexts. A naive implementation of all such contexts would therefore scale exponentially with $K$. However, in a structured chemical realization, the threshold of stage $i$ can instead be formed successively from a base threshold and the evidence generated by previously decoded streams, avoiding explicit instantiation of all branches~\cite{wietfeldChemsicalNetControlling2026}. Following this view, the logical modules can scale linearly with $K$, while the threshold-aggregation complexity grows polynomially with the number of prior decisions~\cite{wietfeldChemsicalNetControlling2026}. Accordingly, DBMC-NOMA is structurally compatible with CRN implementations based on operations such as comparison, translation, and addition. We regard small networks with $K\leq6$ as the most realistic near-term implementation regime. Recent DNA strand-displacement work further supports the feasibility of multi-layer tree-style molecular decision systems, demonstrating cascaded networks beyond 10 layers~\cite{liuDNADecisionTree2025}.

\section{Analytical Bit Error Probability}\label{sec:bep}
We denote the \acs{BEP} of the individual TX$_i$ as $P_{\mathrm{e},i}$ and consider the system \acs{BEP} $P_\mathrm{e,sys}$ as the probability that any transmitted symbol in a given time slot is erroneously detected:
\begin{equation}\label{eq:bep_sys}
    P_{\mathrm{e,sys}} = \frac{1}{K} \sum_{i=1}^{K} P_{\mathrm{e},i}.
\end{equation}
We present the derivation of $P_{\mathrm{e,sys}}$ for a network of $K$ \acp{TX} according to Figure \ref{fig:scenario}. We begin with \ac{DBMC-NOMA} and later also address \ac{TDMA} and \ac{MDMA}. To keep the derivation tractable, the sampling jitter $\Delta_\mathrm{p}$ is disregarded for the analytical evaluation, and therefore, $t_{\mathrm{s},j} = t_{\mathrm{p},j}$.

\subsection{DBMC-NOMA BEP Derivation}
For preparation, we introduce auxiliary variables based on the definitions from Section \ref{sec:system_model}.
Firstly, we write the vector of all average sampling values taken at $t_{\mathrm{p},j}$, for transmission from all \acp{TX} in all time slots as
\begin{multline}
    \mathbf{\Lambda}_j = \left[\lambda_{1,j}[0], \cdots, \lambda_{K,j}[0], \lambda_{1,j}[1], \cdots, \lambda_{K,j}[1],\right. \\ 
    \left.\cdots, \lambda_{1,j}[L], \cdots, \lambda_{K,j}[L]\right].
\end{multline}
Secondly, we define the set of all vectors of length $N$ with binary elements as $\mathbb{B}^{N} = \left\{\left[b_0b_1b_2\cdots b_{N-1}\right] |\, b_i \in \{0,1\} \right\}$. Based on this, we write the vector of the transmitted symbols by all \acp{TX} in all time slots as
\begin{multline}
    \mathbf{S} = [ s_1[0], \cdots, s_K[0], s_1[1], \cdots, s_K[1], \\
    \cdots, s_1[L], \cdots, s_K[L] ]\in \mathbb{B}^{K(L+1)},
\end{multline}
such that 
\begin{multline}
    \mathbf{S}\cdot \mathbf{\Lambda}_j = \sum_{i=1}^{K} \sum_{l=0}^{L} s_i[l]\lambda_{i,j}[l] \\
    = \underbrace{\sum_{\substack{i=1 \\ i \neq j}}^{K} s_i[0]\lambda_{i,j}[0]}_{\mathrm{MAI}} + \underbrace{\sum_{i=1}^{K}\sum_{l=1}^{L}s_i[l]\lambda_{i,j}[l]}_{\mathrm{ISI}},
\end{multline}
as in (\ref{eq:n_rx_noma}).
Similarly, we define a vector of the decoded symbols in the current time slot from TX$_1$ up to TX$_m$
\begin{equation}
    \mathbf{\hat{s}}_m = \left[ \hat{s}_1, \hat{s}_2, \cdots, \hat{s}_m \right] \in \mathbb{B}^{m}.
\end{equation}

We will now consider the case of decoding the symbol from TX$_j$ in the current time slot after having decoded the symbols of TX$_1$ to TX$_{j-1}$. We define the probability that the sample for TX$_j$ after \acs{SIC} is below the threshold \rev{$\tau_j^{\mathbf{\hat{s}}_{j-1}}$} given that TX$_j$ transmitted symbol $s_j[0] = x \in \{0,1\}$ as

\begin{equation}\label{eq:P_j_x}
    P_{j,x} = \mathbb{P}\left( n_{\mathrm{sample},j}^\mathrm{NOMA} < \tau_j^{\mathbf{\hat{s}}_{j-1}} | s_j[0] = x\right).
\end{equation}

Using $\mathcal{P}_\mathrm{CDF}(m;\, \lambda) = \sum_{k=0}^{m} \lambda^k\frac{e^{-\lambda}}{k!}$, which denotes the evaluation of the cumulative density function of the Poisson distribution at $m$, we can calculate the conditional probability associated with $P_{j,x}$ when all transmitted symbols $\mathbf{S}$ and the previously decoded symbols $\mathbf{\hat{s}}_{j-1}$ are given, as
\begin{multline} \label{eq:conditional}
    \mathbb{P}\left(n_{\mathrm{sample},j}<\tau_j^{\mathbf{\hat{s}}_{j-1}}\, |\, s_j[0] = x, \mathbf{S} = \mathbf{S}', \mathbf{\hat{s}}_{j-1} = \mathbf{\hat{s}}_{j-1}'\right)\\
    = \mathcal{P}_\mathrm{CDF}\Biggl(\tau_j^{\mathbf{\hat{s}}_{j-1}'}-1;\, \mathbf{S'}\cdot \mathbf{\Lambda}_j +\lambda_\mathrm{n}\Biggl).
\end{multline}
To arrive at the marginal probability, we must form the sum over all possible cases for $\mathbf{S}$ (with $s_j[0] = x$) and $\mathbf{\hat{s}}_{j-1}$ multiplied by the respective probability of occurrence for each case. Since there are symbols across $L+1$ time slots from $K$ different \acp{TX} considered in the received signal, there are $2^{K(L+1)}$ different equiprobable combinations of transmitted symbols $\mathbf{S}$, which affect the mean of the received signal's Poisson distribution. Additionally, $2^{j-1}$ different possible combinations of detected symbols $\mathbf{\hat{s}}_{j-1}$ for the previously considered \acp{TX} in the current time slot affect the choice of $\tau_j^{\mathbf{\hat{s}}_{j-1}}$ for the \ac{SIC} procedure. The probability of each $\mathbf{\hat{s}}_{j-1}$ occurring depends on $\mathbf{S}$ and on the \acp{BEP} for the previously considered \acp{TX}. Therefore, $\mathbb{P}\left(\mathbf{S} = \mathbf{S}'\in \mathbb{B}^{K(L+1)}\,|\, s_j[0]=x\right)=~\frac{1}{2^{(K(L+1)-1)}}$. 

To calculate the probability of occurrence of the previous $j-1$ decoded symbols, $\mathbf{\hat{s}}_{j-1}$, we write it as the following multiplication of conditional probabilities of a single decoded symbol
\begin{multline}\label{eq:p_hat_marg}
    \mathbb{P}(\mathbf{\hat{s}}_{j-1}=\mathbf{\hat{s}}_{j-1}'| \mathbf{S}=\mathbf{S}', s_j[0]=x) \\
    = \prod_{i=1}^{j-1} \mathbb{P}\left( \hat{s}_{i} = \hat{s}_{i}' | \mathbf{S}=\mathbf{S}', s_j[0]=x, \mathbf{\hat{s}}_{i-1} = \mathbf{\hat{s}}_{i-1}' \right).
\end{multline}
To now calculate the individual factors, we first define
\begin{multline} \label{eq:p_prev}
    P_\mathrm{prev}\Big(i,\mathbf{S}', \mathbf{\hat{s}}_{i-1}'\Big) =\\
     \mathcal{P}_\mathrm{CDF}\Biggl(\tau_i^{\mathbf{\hat{s}}_{i-1}'}-1 ; \mathbf{S}'\cdot \mathbf{\Lambda}_i + \lambda_\mathrm{n}\Biggl).
\end{multline}
The probability of a decoded symbol is then expressed for \rev{two} different possible cases as follows
\rev{\begin{multline} \label{eq:p_hat_cond}
    \mathbb{P}\left( \hat{s}_{i} = \hat{s}_{i}' \,\middle|\,
    \mathbf{S}=\mathbf{S}', s_j[0]=x,
    \mathbf{\hat{s}}_{i-1} = \mathbf{\hat{s}}_{i-1}' \right)\\
    = 
    \begin{cases}
        P_\mathrm{prev}\left(i,\mathbf{S}', \mathbf{\hat{s}}_{i-1}'\right),
        & \hat{s}_i' = 0,\\[1mm]
        1-P_\mathrm{prev}\left(i,\mathbf{S}', \mathbf{\hat{s}}_{i-1}'\right),
        & \hat{s}_i' = 1.
    \end{cases}
\end{multline}}
\rev{Here, \(\mathbf{\hat{s}}_{i-1}'\) denotes the previously decoded SIC
decision vector up to TX$_{i-1}$, whereas the two cases only distinguish between
the current scalar candidate decisions \(\hat{s}_i'\).}
Inserting (\ref{eq:p_prev}) into (\ref{eq:p_hat_cond}), and (\ref{eq:p_hat_cond}) into (\ref{eq:p_hat_marg}) yields the result for $\mathbb{P}\left(\underline{\hat{s}}_{j-1}=\underline{\hat{s}}_{j-1}'\,|\,\mathbf{S}=\mathbf{S}', s_j[0]=x\right)$.
Combining (\ref{eq:p_hat_marg}) with~(\ref{eq:P_j_x}) and (\ref{eq:conditional}), we get
\begin{multline} \label{eq:P_j_x_final}
    P_{j,x} =\!\!\!\!\!\! \sum_{\substack{\mathbf{S}' \in \mathbb{B}^{K(L+1)} \\ s_j[0] = x}}\sum_{\mathbf{\hat{s}}_{j-1}'\in \mathbb{B}^{j-1}} \!\!\!\! \mathbb{P}\left(\mathbf{\hat{s}}_{j-1}=\mathbf{\hat{s}}_{j-1}'| \mathbf{S}=\mathbf{S}',s_j[0]=x\right) \\
    \cdot \frac{1}{2^{(K(L+1)-1)}} \mathcal{P}_\mathrm{CDF}\Biggl(\tau_j^{\mathbf{\hat{s}}_{j-1}'}-1;\, \mathbf{S'}\cdot \mathbf{\Lambda}_j  +\lambda_\mathrm{n}\Biggl). 
\end{multline}
We note that $P_{j,x}$ corresponds to the probability of correct detection if $x=0$, and to the probability of incorrect detection if $x=1$. Therefore, the \ac{BEP} of TX$_j$ is given by
\begin{equation} \label{eq:BEP}
    P_{e,j} = \frac{1}{2}\left(P_{j,1} + (1-P_{j,0})\right).
\end{equation}

Similar but simpler derivations exist for \acs{TDMA}~\cite{rudsariDrugReleaseManagement2019} and \acs{MDMA}, for which we consider $K$ independent communication links~\cite{shiErrorPerformanceAnalysis2017}.

\subsection{SIC Error Propagation Models}
Eq.~(\ref{eq:P_j_x_final}) constitutes a \emph{full} \ac{SIC} model in which the detection of TX$_j$ is averaged over all possible decoded prefixes $\hat{\mathbf{s}}_{j-1}$, weighted by their occurrence probability
$P(\hat{\mathbf{s}}_{j-1}\!\mid \mathbf{S}')$, thereby capturing \ac{SIC} error propagation.

To isolate the impact of error propagation, we additionally consider two reference variants:
(i) \emph{genie-aided \ac{SIC}}, which assumes a perfectly correct prefix $\hat{\mathbf{s}}_{j-1}=\mathbf{s}_{j-1}$ when selecting $\tau_j^{\hat{\mathbf{s}}_{j-1}}$; and
(ii) a conservative \emph{first-error reference}, which evaluates TX$_j$ only under the all-correct prefix and weights it by the probability that all previous \ac{SIC} decisions are correct, i.e.,
$P(\hat{\mathbf{s}}_{j-1}=\mathbf{s}_{j-1}\!\mid \mathbf{S}')$.
As we will show with examples in Section~\ref{sec:ma_comparison}, these references bound the optimistic (no propagation) and pessimistic (catastrophic propagation) behaviors, while the main results use the full model in Eq.~(\ref{eq:P_j_x_final}).

\section{Analytical Evaluation}\label{sec:ma_comparison}

We use the analytical model to evaluate \ac{DBMC-NOMA} under different conditions, analyze the impact of key parameters, and compare \ac{DBMC-NOMA} with \ac{TDMA} and \ac{MDMA}. Table~\ref{tab:analytical_parameters} lists the default parameter values and ranges. The values of distances $d_i$ and \ac{RX} radius $r$ allow the application of the \ac{UCA}, since the condition $r < 0.15 d_i$ is valid and the channel is considered sufficiently long compared to the radius.

\subsection{Performance Metrics}
To measure the system's performance, we will utilize two metrics depending on the evaluated scenario and context.
\subsubsection{System bit-error probability (BEP)} The system \ac{BEP}, $P_\mathrm{e,sys}$, as described in Section~\ref{sec:bep} and Eq.~(\ref{eq:bep_sys}) will be used to evaluate the performance between different scenarios that use the same MA scheme. Specifically, we will use it to investigate the effects of parameters on the performance of \ac{DBMC-NOMA}.
\subsubsection{Mutual Information (MI)} The \ac{MI}~\cite{coverElementsInformationTheory1991} of the induced binary-input/binary-output channel after threshold detection can be derived from the probability distributions of transmitted and received symbols and represents the amount of bits of information that can be conveyed from \ac{TX} to \ac{RX}. \Ac{MI} due to a transmission from TX$_i$ in the current time slot is denoted as $\mathcal{I}_i$. Here, we directly reuse the notation from Section~\ref{sec:bep}, i.e., the transmitted bit in the current time slot is $s_i[0]\in\{0,1\}$ and the corresponding hard decision is $\hat{s}_i\in\{0,1\}$.

Using the analytical probabilities $P_{i,0}$ and $P_{i,1}$ from Section~\ref{sec:bep}, the transition probabilities of the induced binary channel are
\begin{align}
    p_{00}^{(i)} \triangleq \mathbb{P}(\hat{s}_i=0\mid s_i[0]=0) = P_{i,0},\\
    p_{10}^{(i)} \triangleq \mathbb{P}(\hat{s}_i=1\mid s_i[0]=0) = 1-P_{i,0},\\
    p_{01}^{(i)} \triangleq \mathbb{P}(\hat{s}_i=0\mid s_i[0]=1) = P_{i,1},\\
    p_{11}^{(i)} \triangleq \mathbb{P}(\hat{s}_i=1\mid s_i[0]=1) = 1-P_{i,1}.
\end{align}
Assuming equiprobable symbols $\mathbb{P}(s_i[0]=0)=\mathbb{P}(s_i[0]=1)=\frac{1}{2}$ as in the analytical evaluation in Section~\ref{sec:bep}, the post-detection MI for TX$_i$ is
\begin{multline}
    \mathcal{I}_i \triangleq I(s_i[0];\hat{s}_i)
    = \sum_{x\in\{0,1\}}\sum_{\hat{x}\in\{0,1\}} \mathbb{P}(s_i[0]=x)\\ \cdot \mathbb{P}(\hat{s}_i=\hat{x}\mid s_i[0]=x)\log_2\!\frac{\mathbb{P}(\hat{s}_i=\hat{x}\mid s_i[0]=x)}{\mathbb{P}(\hat{s}_i=\hat{x})},
\end{multline}
where $\mathbb{P}(\hat{s}_i=\hat{x})=\sum_{x\in\{0,1\}}\mathbb{P}(s_i[0]=x)\,\mathbb{P}(\hat{s}_i=\hat{x}\mid s_i[0]=x)$.
For \ac{MDMA} and \ac{NOMA} all \acp{TX} transmit in all time slots. In contrast, the \ac{MI} for \ac{TDMA} must be averaged across the $K$ \acp{TX} since one \ac{TX} transmits at a time. We use $\mathcal{I}_\mathrm{sys} = \sum_{i=1}^{K}\mathcal{I}_i$ as a throughput proxy per time slot for \ac{MDMA} and \ac{NOMA}, and $\mathcal{I}_\mathrm{sys} = \frac{1}{K}\sum_{i=1}^{K}\mathcal{I}_i$ for \ac{TDMA}. Therefore, we will utilize \ac{MI} particularly for the comparison between the different MA schemes. 
Both $P_{\mathrm e}$ and $\mathcal{I}$ are computed from the same induced binary-input/binary-output channel after threshold detection. Hence, reducing $P_{\mathrm e}$ typically increases $\mathcal{I}$, although the mapping is not strictly one-to-one for asymmetric error events (e.g., $p_{01}\neq p_{10}$).
Importantly, the \ac{MI} comparisons in this section focus on the \emph{data phase} and assume optimally chosen parameters (via exhaustive search) to benchmark the \ac{MA} schemes, whereas the DBMC-aNOMAly evaluation in Section~\ref{sec:protocol_eval} targets \emph{runtime} parameter adaptation under unknown and time-varying conditions using pilot symbols.
We will determine and discuss the impact of overhead on the net \ac{MI} in Section~\ref{sec:protocol_eval}.

Additionally, the \ac{SNR} is utilized as a measure of relative noise level and is defined as the ratio between the expected signal $\tilde\lambda_i$ and the additive noise for each TX$_i$
    \begin{equation}
        \mathrm{SNR} = 20\log_{10}\frac{\tilde\lambda_i}{\lambda_{\mathrm{n}}}.
    \end{equation}

To validate the analytically derived results with another method, \rev{Monte Carlo simulations} were conducted for matching scenarios. Here, the Poisson distributions of the received signals were sampled for a large number of randomly generated symbol vectors and the different \ac{MA} schemes were applied. Eq.~(\ref{eq:P_j_x_final}) is algebraically equivalent to an exact \emph{direct enumeration} of all $2^{K(L+1)}$ transmitted symbol vectors and all $2^{j-1}$ SIC decision prefixes for stream $j$, weighting each branch by its occurrence probability. The \rev{Monte Carlo simulations} presented alongside the analytical results serve as an independent numerical check by approximating the same expectation via random sampling of symbol vectors and Poisson observations.

\begin{table}[t]
\caption{Parameters for the Analytical Evaluation}
\label{tab:analytical_parameters}
\resizebox{\columnwidth}{!}{%
\begin{tabular}{lll}
\toprule
\textbf{Parameter}    & \textbf{Symbol}                            & \textbf{Values ($\underline{\mathrm{Default}}$)}                          \\ \midrule
Number of TXs            & $K$                                        & $\underline{2}, 3, 4, 5, 6$                                     \\
TX distances          & $d_i$                 & $\underline{10\,\qty{}{\micro\meter}}, 14\,\qty{}{\micro\meter}$                \\
RX radius             & $r$                                        & $\qty{1}{\micro\meter}$                 \\
Diffusion coefficient & $D$                                        & $\qty{e-9}{\meter\squared\per\second}$ \\
Symbol period           & $T$           & $1\ \unit{\second}$\\
\ac{ISI} symbols       & $L$            & $1$\\
Signaling-molecule-to-noise ratio                   & SNR & $\{[-50, 30], \underline{\infty}\}\, \unit{\decibel}$                                  \\ 
Molecule budget per \acs{TX}                   & $N_\mathrm{TX,max}$ & $[10^5, \underline{10^6}]$ molecules                                  \\
\bottomrule
\end{tabular}%
}
%\vspace{-0.5cm}
\end{table}

\subsection{Parameter Analysis for DBMC-NOMA}

This section focuses on specific parameters of \ac{DBMC-NOMA} to highlight isolated effects on the performance.

\begin{figure}[t]
    \centering
    \includegraphics[width=0.8\linewidth]{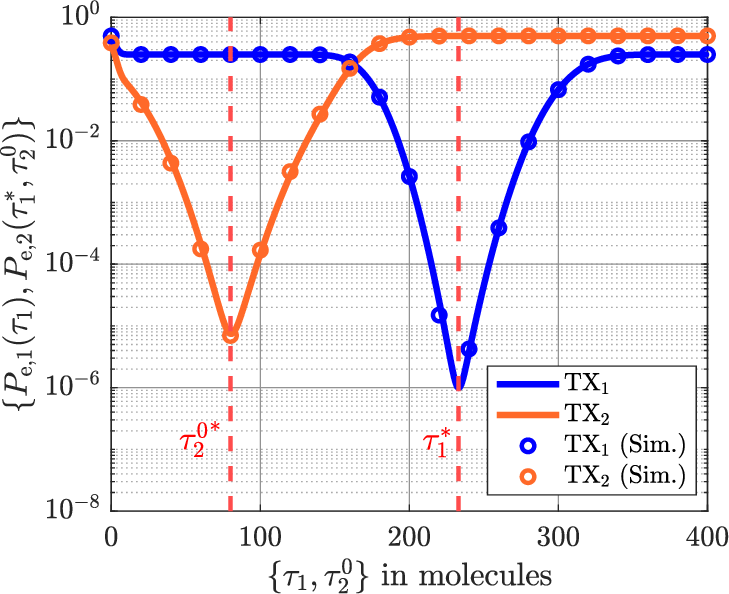}
    \caption{Analytical \acp{BEP} $P_{\mathrm{e},i}$ versus detection thresholds for a \ac{DBMC-NOMA} system with $K=2$ \acp{TX}, with simulation markers for validation. The plot shows $P_\mathrm{e,1}(\tau_1)$, independent of TX$_2$, and $P_\mathrm{e,2}(\tau_1^*, \tau_2^0)$ for optimal $\tau_1^*$. Here, $N_\mathrm{TX,1} = 10^6$, $N_\mathrm{TX,2}\approx 0.5\cdot 10^6$, and $d_1 = d_2 = \qty{10}{\micro\meter}$; all other parameters follow Table~\ref{tab:analytical_parameters}.}
    \label{fig:opt_threshold}
\end{figure}

\begin{figure}[t]
    \centering
    \includegraphics[width=0.8\linewidth]{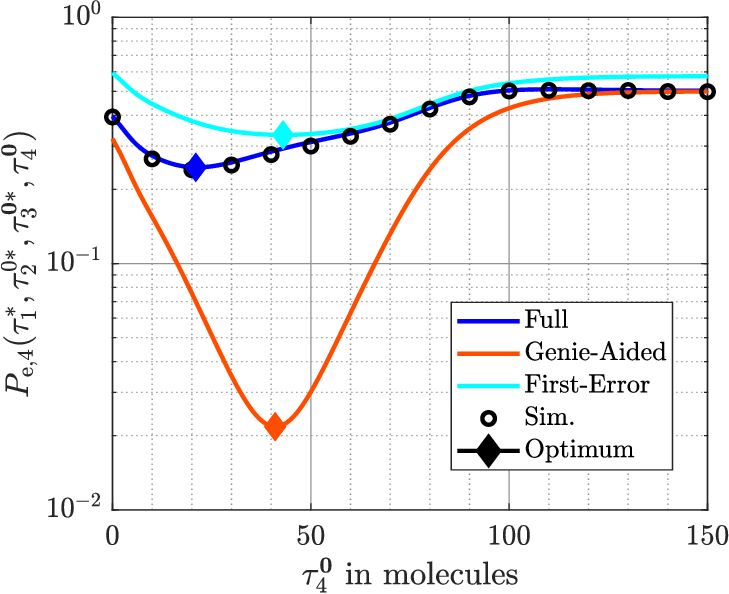}
    \caption{SIC error-propagation models for downstream threshold design in a representative \ac{DBMC-NOMA} system with $K=4$ \acp{TX}, $d_i=\qty{10}{\micro\meter}\ \forall i$, and $N_\mathrm{TX,1}=10^6$, $N_\mathrm{TX,2}=0.5\cdot10^6$, $N_\mathrm{TX,3}=0.33\cdot10^6$, $N_\mathrm{TX,4}=0.25\cdot10^6$. The full model from Eq.~(\ref{eq:P_j_x_final}) is compared to the genie-aided and first-error reference models from Section~\ref{sec:bep}; $\tau_1^*$, $\tau_2^*$, and $\tau_3^*$ are fixed, while $\tau_4^\mathbf{0}$ is swept. Simulation markers for the actual SIC detector closely match the full analytical model. All other parameters follow Table~\ref{tab:analytical_parameters}.}
    \label{fig:sic_modes}
\end{figure}

\subsubsection{Optimizing the Detection Threshold}

In Fig.~\ref{fig:opt_threshold}, $P_{\mathrm{e},i}$ is plotted over the respective $\tau_1$ and $\tau_2^0$ for a system with $K=2$ \acp{TX}. For ease of demonstration in the plot, it is assumed that $\tau_2^1 = \tau_2^0 + \tilde{\lambda}_1$, mirroring the classical \ac{SIC} scheme in Fig.~\ref{fig:sic_comparison}. The number of transmitted molecules is chosen as $N_\mathrm{TX,1} = 10^6$ and $N_\mathrm{TX,2} \approx 0.5\cdot10^6$. All other parameters correspond to the default values in Table~\ref{tab:analytical_parameters}.
The optimization process of the thresholds for \ac{DBMC-NOMA} can be conducted iteratively, similar to the structure of the \ac{SIC} procedure. The optimum threshold $\tau_1^*$ can be chosen independently of the thresholds for TX$_2$. Then, the choice of $\tau_2^0$ and $\tau_2^1$ depends on the resulting $P_\mathrm{e,1}^*$.
In Fig.~\ref{fig:opt_threshold}, $P_\mathrm{e,2}$ is shown for $\tau_1 = \tau_1^*$ and therefore, a fixed $P_\mathrm{e,1}$. Subsequently, we can choose ${\tau_2^0}^*$ at the minimum of the orange line plot.
The figure shows how we can identify a unique optimum value for the detection thresholds in the \ac{DBMC-NOMA} system. This procedure could be continued similarly for network sizes $K>2$ using the same principle, as the thresholds of TX$_i$ are always independent of any TX$_j$ with $j>i$.

To clarify the role of \ac{SIC} error propagation in the threshold design, Fig.~\ref{fig:sic_modes} compares the full SIC model of Eq.~(\ref{eq:P_j_x_final}) to the genie-aided and first-error reference models introduced in Section~\ref{sec:bep}. For a representative $K=4$ scenario with equal \ac{TX}-\ac{RX} distances and unequal emitted molecule counts $N_{\mathrm{TX},i}$, the upstream thresholds are fixed to the full-model optimum and only the downstream threshold $\tau_4^\mathbf{0}$ is varied. The \rev{Monte Carlo simulation} markers closely follow the analytical full-\ac{SIC} curve, confirming that the full model captures the behavior of the actual SIC detector. At the same time, the minima of the three curves are visibly different. We have chosen $K=4$, as the differences between the models grow as the number of \acp{TX} increases. The averaging over all decoded symbols in Eq.~(\ref{eq:P_j_x_final}) incorporates error propagation explicitly, while the genie-aided and first-error cases serve as optimistic and conservative reference models.

\subsubsection{Optimizing the Emitted Number of Molecules}\label{subsubsec:optNumMol}

\begin{figure}[t]
    \centering
    \includegraphics[width=0.8\linewidth]{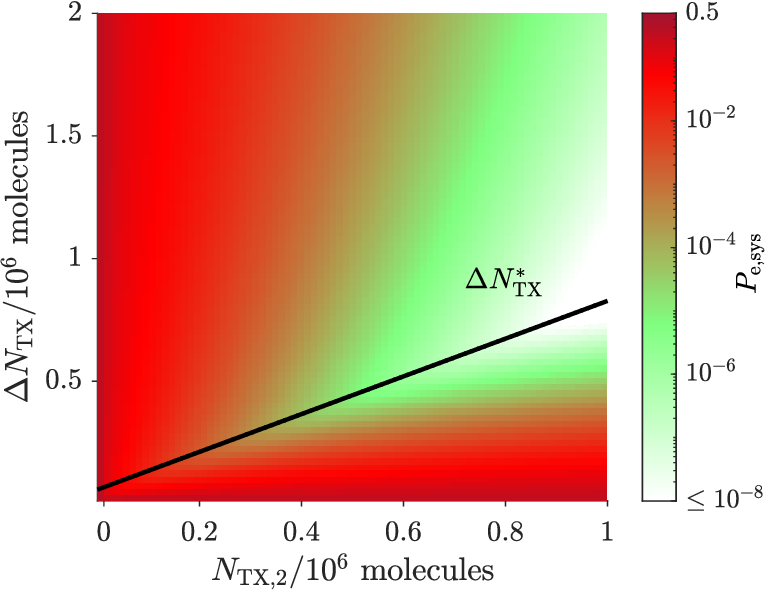}
    \caption{\Ac{DBMC-NOMA} system \ac{BEP} $P_\mathrm{e,sys}$ depicted as a color-coded heatmap with $K=2$ \acp{TX}. Number of emitted molecules of TX$_2$, $N_\mathrm{TX,2}$, on the $x$-axis, and the difference $\Delta N_\mathrm{TX} = N_\mathrm{TX,1} - N_\mathrm{TX,2}$ on the $y$-axis. For each point, detection thresholds are chosen optimally. Black line indicates the respective optimum value of $\Delta N_\mathrm{TX}$ w.r.t $P_\mathrm{e,sys}$ for each value of $N_\mathrm{TX,2}$. All other parameters according to Table~\ref{tab:analytical_parameters}.}
    % \caption{Demonstrate finding optimum numMol with 2 TX example heatmap, for optimum thresholds, identifying optimum ratio between numMol. For sync case: BER for varying numMol for TX1 and TX2, showing distinct optimum line.}
    \label{fig:opt_numMol}
\end{figure}

Fig.~\ref{fig:opt_numMol} depicts a heatmap of the system \ac{BEP} $P_\mathrm{e,sys}$ over a varying $N_\mathrm{TX,2}$ on the $x$-axis as well as $\Delta N_\mathrm{TX} = N_\mathrm{TX,1} - N_\mathrm{TX,2} > 0$ on the $y$-axis. The network has $K=2$ \acp{TX}. For every point on the map, the optimum thresholds have been chosen via exhaustive search for the specific combination of $N_\mathrm{TX,1}, N_\mathrm{TX,2}$.
On the graph, we can identify an optimum value $\Delta N_\mathrm{TX}^*$ for each $N_\mathrm{TX,2}$ forming a line.

This is caused by the application of the \ac{SIC} procedure. If $\Delta N_\mathrm{TX}$ was reduced from the optimum, the separation of signal contributions from each \ac{TX} would decrease, making it more difficult to choose an appropriate threshold for TX$_1$, independent of the transmission by TX$_2$. This is only possible for sufficient difference between the two contributions. $P_\mathrm{e,1}$ would increase and, therefore, also the detector is more likely to apply the wrong threshold for TX$_2$, increasing $P_\mathrm{e,2}$ as well.
The standard deviation of the signal in the Poisson model is equal to the square-root of its mean, $\sigma = \sqrt\lambda$. As a result, if $\Delta N_\mathrm{TX}$ is increased above the optimum, the signal standard deviation at the \ac{RX} will increase together with the total signal mean. For TX$_1$, the relative value $\frac{\sigma}{\lambda} = \frac{1}{\sqrt\lambda}$ will decrease, since the mean keeps rising with it. However, the signal standard deviation for TX$_2$ will increase more and more, leading to an increase in $P_\mathrm{e,2}$.

Following the optimum $\Delta N_\mathrm{TX}^*$ line from left to right, we also observe that lower and lower values of $P_\mathrm{e,sys}$ are achieved. This is also due to the signal-dependent value of the standard deviation. In this case, however, both $N_\mathrm{TX,1}$ and $N_\mathrm{TX,2}$ are increased. Therefore, the standard deviation relative to the mean decreases for both \acp{TX}, the detection of both samples is more reliable, and $P_\mathrm{e,sys}$ decreases.
Consequently, the color-coded low-error region widens visually as the minimum $P_\mathrm{e,sys}$ drops and the surface around $\Delta N_\mathrm{TX}^*$ becomes flatter and less sensitive to small deviations.

\begin{figure}[t]
    \begin{subfigure}{\linewidth}
    \centering
        \includegraphics[width=0.8\linewidth]{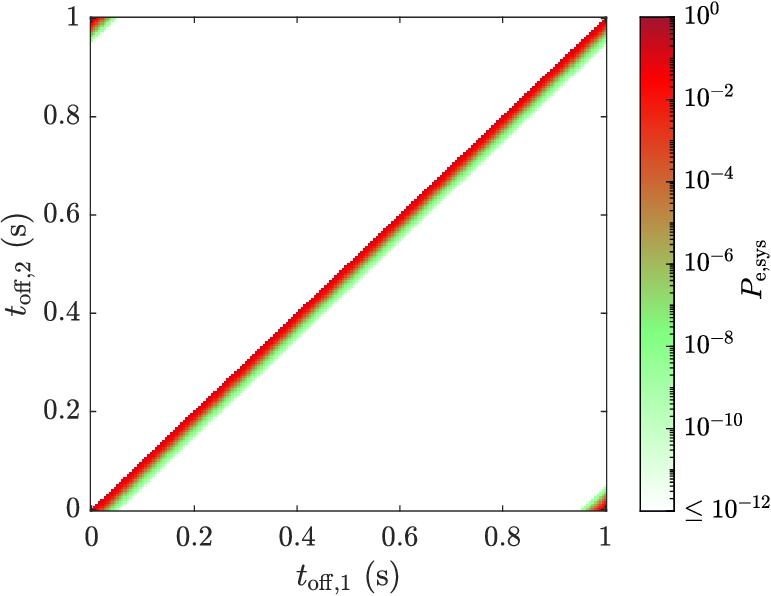}
        \caption{$d_1=d_2=\qty{10}{\micro\meter}$}
        \label{fig:opt_offset:subfig:sameDist}
    \end{subfigure}
    \hfill
    \begin{subfigure}{\linewidth}
    \centering
        \includegraphics[width=0.8\linewidth]{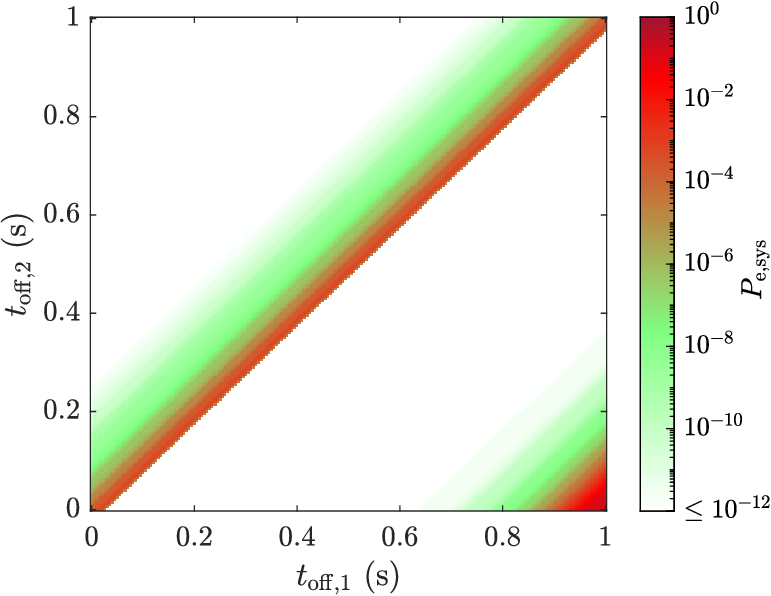}
        \caption{$d_1 = \qty{10}{\micro\meter},\,d_2=\qty{14}{\micro\meter}$}
        \label{fig:opt_offset:subfig:diffDist}
    \end{subfigure}
    \caption{\Ac{DBMC-NOMA} system \ac{BEP}, $P_\mathrm{e,sys}$, depicted as a color-coded heatmap with the synchronization offsets of TX$_1$, $t_\mathrm{off,1}$, and TX$_2$, $t_\mathrm{off,2}$, on the $x$- and $y$-axis, respectively. For all points, the detection thresholds were chosen optimally. All other parameters according to Table~\ref{tab:analytical_parameters}.}
    \label{fig:opt_offset}
\end{figure}

\subsubsection{Varying the Time Offset}

In previous work on \ac{DBMC-NOMA}~\cite{wietfeldDBMCNOMAEvaluatingNOMA2024, wietfeldErrorProbabilityOptimization2024c}, only the detection thresholds and the number of emitted molecules were considered as variable parameters in the system to optimize the performance. In this paper, we will also consider the choice of time offset $t_{\mathrm{off},j}$ between different \acp{TX}, as defined in Section~\ref{sec:system_model}. In the previous investigations, the assumption was that the system is fully synchronized and $t_{\mathrm{off},j} = 0\ \forall_j$.
Relaxing this assumption, Fig.~\ref{fig:opt_offset} depicts heatmaps of $P_\mathrm{e,sys}$ for a system of 2 \acp{TX} and a varying $t_{\mathrm{off},1}$ and $t_{\mathrm{off},2}$ on the $x$-axis and $y$-axis, respectively. Similarly to the case discussed in Section~\ref{subsubsec:optNumMol}, the optimum thresholds were identified via exhaustive search and chosen for each point on the heatmap. Additionally, $N_\mathrm{TX,1} = N_\mathrm{TX,2} = 10^6$.

The plot in Fig.~\ref{fig:opt_offset:subfig:sameDist} shows limited areas with a high error probability, which we will denote as \emph{worst-case offset} areas. These correspond to cases with either $t_{\mathrm{off},1} \approx t_{\mathrm{off},2}$ or $t_{\mathrm{off},j}\approx T=1$.
Here, the peaks of the arriving channel impulse responses align, leading to significant \ac{MAI}, as calculated in Eq.~\ref{eq:n_rx_noma} in Section~\ref{sec:system_model}. Despite the long tail of the \ac{DBMC} channel impulse response, the peak itself is relatively sharp.
Therefore, next to the worst-case offset areas, there is a steep decline in \ac{MAI} and drop-off in $P_\mathrm{e,sys}$.
This suggests that there is a way of optimizing performance by avoiding the worst-case offset areas and simultaneously optimizing the detection thresholds. We will address this later with the \acf{WCAM} in Section~\ref{sec:protocol}.
\begin{figure}[!t]
    \begin{subfigure}{\linewidth}
    \centering
        \includegraphics[width=0.8\linewidth]{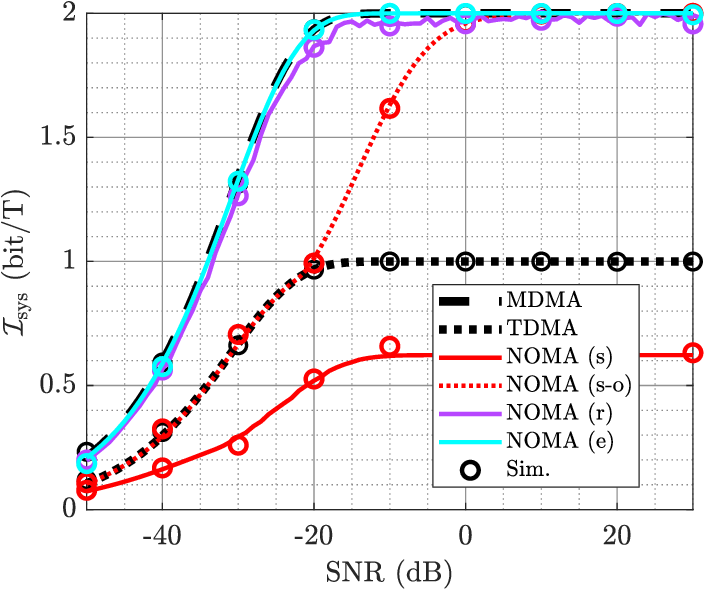}
        \vspace{-0.1cm}
        \caption{$N_\mathrm{TX,max} = 10^6$, $d_1=d_2=\qty{10}{\micro\meter}$.}
        \label{fig:analytical_snr:subfig:1e6}
    \end{subfigure}
    \hfill
    \begin{subfigure}{\linewidth}
    \centering
        \includegraphics[width=0.8\linewidth]{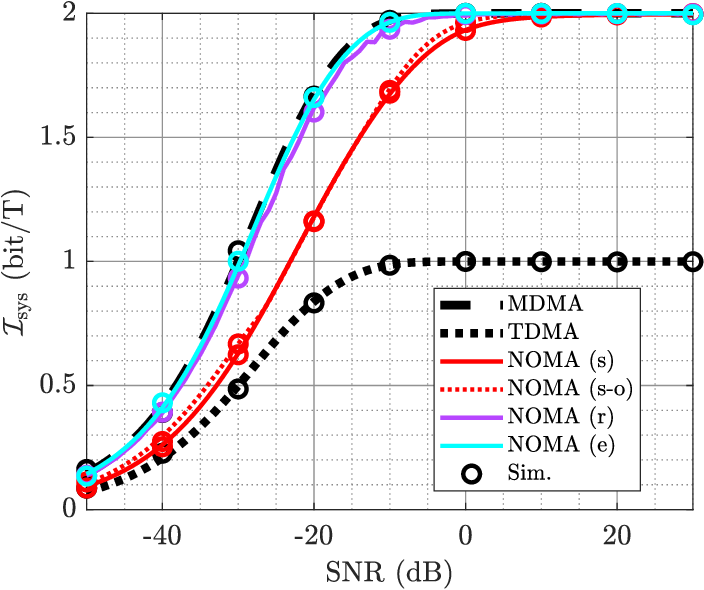}
        \vspace{-0.1cm}
        \caption{$N_\mathrm{TX,max} = 10^6$, $d_1 = \qty{10}{\micro\meter},\,d_2=\qty{14}{\micro\meter}$.}
        \label{fig:analytical_snr:subfig:1e6_diffDist}
    \end{subfigure}
    \hfill
    \begin{subfigure}{\linewidth}
    \centering
        \includegraphics[width=0.8\linewidth]{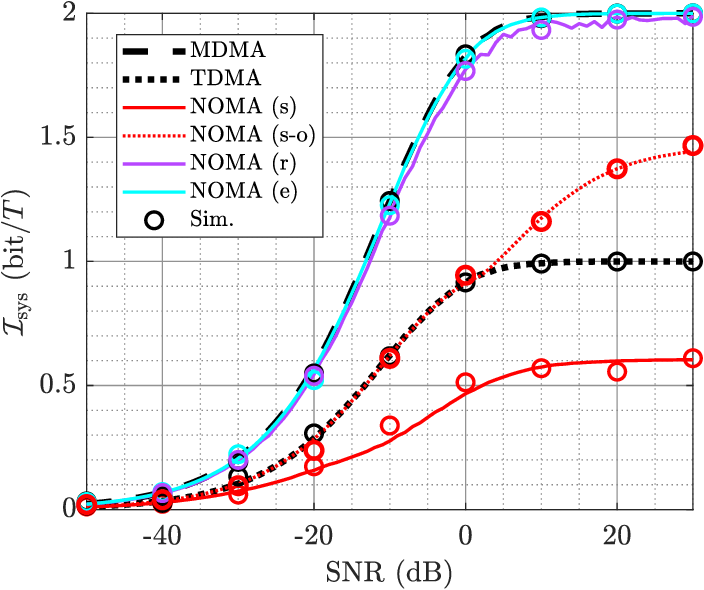}
        \vspace{-0.1cm}
        \caption{$N_\mathrm{TX,max} = 10^5$, $d_1=d_2=\qty{10}{\micro\meter}$}
        \label{fig:analytical_snr:subfig:1e5}
    \end{subfigure}
    \caption{Mutual information $\mathcal{I}_\mathrm{sys}$ per symbol period $T$ over the SNR for three different \ac{MA} schemes, and four different variants of \ac{DBMC-NOMA}. Synchronized \emph{(s)}, synchronized and $N_\mathrm{TX}$ optimized \emph{(s-o)}, random offsets \emph{(r)}, evenly distributed offsets \emph{(e)}. Analytical results and corresponding \rev{Monte Carlo simulations} are depicted for validation. Detection thresholds are chosen optimally. Results are shown for two different values of the molecule budget per \ac{TX}, $N_\mathrm{TX,max}$. For all other parameters see Table~\ref{tab:analytical_parameters}.}
    \label{fig:analytical_snr}
\end{figure}
To assess the sensitivity of these worst-case offset areas to unequal distances, Fig.~\ref{fig:opt_offset:subfig:diffDist} shows an example for $d_1=\qty{10}{\micro\meter}$ and $d_2=\qty{14}{\micro\meter}$. The unequal-distance case exhibits the same qualitative worst-case structure as the equal-distance case, i.e. a high-error region around approximately equal effective offsets, while this region becomes broader and less localized. This is due to the lower signal arriving from the farther \ac{TX}$_2$, which makes it more sensitive to interference. Hence, the worst-case-offset concept remains applicable beyond the equal-distance setting, although the quantitative sensitivity to offsets differs with distance asymmetry.

\subsection{Comparison of MA Schemes}

In this subsection, we will compare the performance of \ac{TDMA}, \ac{MDMA}, and \ac{DBMC-NOMA} under different conditions. The parameters utilized throughout this evaluation are listed in Table~\ref{tab:analytical_parameters} with highlighted default values that are used in the absence of a separate definition. We note that \ac{MDMA} acts as an upper-bound target for the performance, as it allows for $K$ entirely independent channels, the performance of which cannot be surpassed by another \ac{MA} scheme. As described in Section~\ref{sec:system_model}, the implementation of \ac{MDMA} would entail an increase in system complexity, related to the increased number of different molecule types. Again, the results include the analytical \ac{BEP} evaluation derived in Section~\ref{sec:bep} and \rev{Monte Carlo simulations} for validation.

\subsubsection{Offset Cases}
Four different configurations of \ac{DBMC-NOMA} will be considered during the evaluation. In all cases, we choose the optimum detection thresholds via exhaustive search.
\begin{itemize}
    \item Synchronized \emph{(s)}: All \acp{TX} send at the same time, i.e. $t_{\mathrm{off},i} = 0\ \forall_i$. Represents the assumed \emph{worst-case} scenario.
    \item Synchronized with optimized number of emitted molecules \emph{(s-o)}: $t_{\mathrm{off},i}=0\ \forall_i$ and we choose the optimum $N_{\mathrm{TX},i}$ via exhaustive search. This is the same setup used in previous work~\cite{wietfeldDBMCNOMAEvaluatingNOMA2024, wietfeldErrorProbabilityOptimization2024c}.
    \item Random offset \emph{(r)}: $t_{\mathrm{off},i}$ are chosen from a random uniform distribution $\mathcal{U}[0,T]$ and the results are averaged over 200 samples. Represents the assumed \emph{average-case} scenario.
    \item Even offset \emph{(e)}: $t_{\mathrm{off},i}$ are distributed evenly across the range $[0,T]$. Represents the assumed \emph{best-case} scenario.
\end{itemize}
If $N_{\mathrm{TX},i}$ is optimized, we assume a maximum molecule budget per \ac{TX} of $N_\mathrm{TX,max}$, up to which each $N_{\mathrm{TX},i}$ can be varied. For all other cases, we assume $N_{\mathrm{TX},i} = N_\mathrm{TX,max}\ \forall_i$.

Preliminary results have shown and the mathematical formulas in Section~\ref{sec:system_model} suggest that \ac{TDMA} and \ac{MDMA} are not or only marginally affected by the different offset configurations, due to the lack of \ac{MAI}. Therefore, we show only one curve for these two schemes.

\subsubsection{Varying Noise Level}

Fig.~\ref{fig:analytical_snr} shows the mutual information per time slot of the entire system, $\mathcal{I}_\mathrm{sys}$, on the $y$-axis, and the SNR, i.e. the additive noise level, on the $x$-axis. The results are shown for $K=2$ and two different values of $N_\mathrm{TX,max}$. In addition to the default equal-distance setup, we include one exemplary unequal-distance case to assess the robustness of the observed trends. We will start with Fig.~\ref{fig:analytical_snr:subfig:1e6}, where $N_\mathrm{TX,max} = 10^6$ and distances are equal.
As expected, \ac{MDMA} is the upper bound for all schemes. The performance of the different \ac{DBMC-NOMA} configurations varies significantly.
In the (s) case, it performs worse than all schemes including \ac{TDMA}, due to \ac{MAI} and the same value of $N_{\mathrm{TX},i}$.
For the (s-o) case, the \ac{MAI} is managed more effectively, i.e. $N_\mathrm{TX,2}$ is reduced as needed. We can observe that $\mathcal{I}_\mathrm{sys}^\mathrm{NOMA}\approx\mathcal{I}_\mathrm{sys}^\mathrm{TDMA}$ for low SNR, and $\mathcal{I}_\mathrm{sys}^\mathrm{NOMA}\approx\mathcal{I}_\mathrm{sys}^\mathrm{MDMA} \approx 2$ bit/$T$ for high SNR. 
However, in both asynchronous cases (r) and (e), \ac{DBMC-NOMA} always outperforms \ac{TDMA} and achieves $\mathcal{I}_\mathrm{sys}^\mathrm{NOMA}\approx\mathcal{I}_\mathrm{sys}^\mathrm{MDMA}$ for all SNR values. It is visible that the randomized (r) case underperforms (e). This is due to the unavoidable inclusion of some worst-case offset samples in the averaged result for (r), in which the performance more closely resembles (s).

Fig.~\ref{fig:analytical_snr:subfig:1e6_diffDist} provides an exemplary unequal-distance cross-check for $d_1=\qty{10}{\micro\meter}$ and $d_2=\qty{14}{\micro\meter}$. We observe that the qualitative ordering of the considered schemes and \ac{DBMC-NOMA} variants remains unchanged: the synchronous case (s) remains the weakest, while the asynchronous cases (r) and (e) still outperform \ac{TDMA} and approach \ac{MDMA} at moderate and high SNR. At the same time, the gap between the asynchronous and synchronized cases becomes smaller than in Fig.~\ref{fig:analytical_snr:subfig:1e6}, indicating that unequal distances mainly affect the quantitative gain, but not the overall conclusion that offset diversity continues to provide a performance benefit in this unequal-distance cross-check.

Looking at Fig.~\ref{fig:analytical_snr:subfig:1e5}, where $N_\mathrm{TX,max} = 10^5$, we see a similar behavior with only slight differences.
A major disadvantage of the $N_\mathrm{TX}$ optimization becomes apparent, as the (s-o) case does not reach \ac{MDMA} performance even for very large SNR values.

\subsubsection{Varying Network Size}

\begin{figure}[t]
    \centering
    \includegraphics[width=0.8\linewidth]{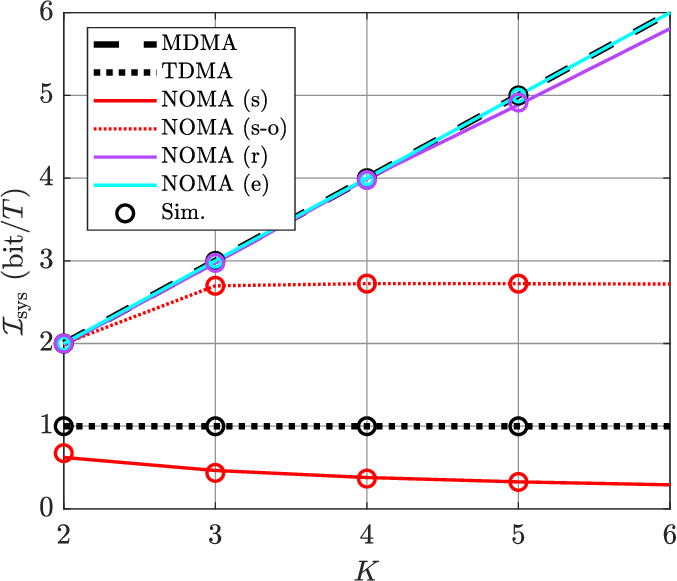}
    \caption{Mutual information $\mathcal{I}_\mathrm{sys}$ per symbol period $T$ over the number of \acp{TX}, $K$, for three different \ac{MA} schemes, and four different variants of \ac{DBMC-NOMA}. Synchronized \emph{(s)}, synchronized and $N_\mathrm{TX}$ optimized \emph{(s-o)}, random offsets \emph{(r)}, evenly distributed offsets \emph{(e)}. Analytical results and corresponding \rev{Monte Carlo simulations} are depicted for validation. All other parameters according to Table~\ref{tab:analytical_parameters}.}
    \label{fig:analytical_numTx}
\end{figure}

The network size $K$ is varied on the $x$-axis with $\mathcal{I}_\mathrm{sys}$ on the $y$-axis in Fig.~\ref{fig:analytical_numTx}.
We can observe several crucial effects. Firstly, in the synchronous case (s), the system performance deteriorates with a growing number of \acp{TX}, meaning that existing \acp{TX} are negatively affected to a larger extent than the added throughput generated by the new \acp{TX}. This is in line with the expected effects of significant \ac{MAI}.
Secondly, the management of $N_\mathrm{TX}$ in the (s-o) case avoids the performance deterioration, but leads to a plateau after a certain network size $K$ is reached. Any added \ac{TX}$_i$ will then always be assigned $N_{\mathrm{TX},i}=0$ after this point, leaving the performance of previous \acp{TX} unaffected.
Similar to the observations when varying the SNR in Fig.~\ref{fig:analytical_snr}, \ac{DBMC-NOMA} in the (e) configuration matches the performance of \ac{MDMA} for all considered network sizes. In the (r) case, system performance slightly falls short of the upper bound, with the difference growing for higher values of $K$, as the probability of hitting a worst-case offset sample increases and the performance impact of the (s) case grows simultaneously.

\subsection{Summary of Analytical Evaluation}

From the results above, we can take away the following main points:
\begin{enumerate}
    \item Allowing for the synchronization offset between the \acp{TX} as a third variable promises significant performance benefits.
    \item These performance benefits appear to be larger than the ones provided by the optimization of $N_{\mathrm{TX},i}$, which we will see later involves a large amount of effort.
    \item It appears that the full potential performance can be achieved, if the few worst-case offset cases can be avoided to close the gap between the (r) and (e) configuration.
\end{enumerate}

The above ideal results, particularly for the evenly distributed offsets (e), assume optimal thresholds and operation away from worst-case offset configurations. In practice, offsets and effective channel conditions are generally unknown and time-varying, motivating the DBMC-aNOMAly protocol introduced in the next section as a runtime mechanism for threshold adaptation and worst-case-offset avoidance.

\section{BEP Minimization Protocol}\label{sec:protocol}

Having benchmarked the achievable data-phase performance of the considered \ac{MA} schemes under optimized parameters, we now turn to the practical question of how a low-complexity protocol can approach favorable operating points under unknown and changing conditions.

We have seen that the detection thresholds, the number of emitted molecules, and the synchronization offset are the major controllable parameters that influence the performance of a \ac{DBMC-NOMA} system. 
Therefore, the following optimization problem can be posed
    \begin{align}
            \left\{\mathcal{T}_i^*, t_{\mathrm{off},i}^*, N_{\mathrm{TX},i}^*\right\}_{i=1}^{K} & = \argmin_{\left\{\mathcal{T}_i, t_{\mathrm{off},i}, N_{\mathrm{TX},i}\right\}_{i=1}^{K}} P_\mathrm{e,sys} \\
            \mathrm{s.t.}\ & N_{\mathrm{TX},i} \leq N_\mathrm{TX,max} \forall i\in {1,\dots, K}. \nonumber
    \end{align}
We have derived the calculation of the \ac{BEP} $P_\mathrm{e,sys}$ in Section~\ref{sec:bep}. Analytical solutions, global optimization algorithms~\cite{chouhanOptimalTransmittedMolecules2019, chengOptimizationDecisionThresholds2022}, and data-driven methods can often find optimal values~\cite{wietfeldErrorProbabilityOptimization2024c}. However, as discussed in Section~\ref{sec:introduction}, they require channel knowledge, accurate evaluation of complex functions and derivatives, or computing power beyond the limited capabilities of future nanoscale nodes.

In previous work, we have proposed and evaluated a simple greedy heuristic based on pilot symbols and have shown that it reliably attains the optimum values for $\mathcal{T}_i$ and $N_{\mathrm{TX},i}$~\cite{wietfeldErrorProbabilityOptimization2024c}. In a different work, we showed how simple operations like comparisons and additions for a \ac{DBMC-NOMA} scheme can be modeled and simulated as a \ac{CRN} framework~\cite{wietfeldChemSICalEvaluatingStochastic2025}. The optimization heuristic is created using similar operations and could also be modeled as a \ac{CRN}.
Based on those results, we propose an extended pilot-symbol-based protocol, \ac{DBMC-aNOMAly}, that improves upon multiple shortcomings of the initial proposal. Firstly, it applies to the asynchronous system defined in Section~\ref{sec:system_model}, removing the need for a high-effort synchronization of the entire network. Secondly, we design a very simple heuristic to avoid the worst-case offset cases, as observed in Section~\ref{sec:ma_comparison}, and we show that the emitted-molecule-count optimization can be made unnecessary. The latter relied on a complex feedback mechanism between \ac{RX} and \acp{TX} that could be prone to errors~\cite{wietfeldErrorProbabilityOptimization2024c}. Also, the optimization of the emitted number of molecules can only be defined in a straightforward manner for a system with $K=2$.

We first describe the mechanisms for dealing with detection thresholds $\mathcal{T}_i$, synchronization offsets $t_{\mathrm{off},i}$, and emitted number of molecules $N_{\mathrm{TX},i}$ separately, and then combine them to form the \ac{DBMC-aNOMAly} protocol.

\subsection{Optimizing the Detection Thresholds}

\begin{algorithm}[t]
\caption{Detection Threshold Optimization Algorithm}\label{alg:threshold}
\begin{algorithmic}[0]
\footnotesize

\State \textsc{Input:} $\mathcal{T}_i^\mathrm{NOMA} = \{\tau_i^\mathbf{\tilde s_{i-1}}\}\ \forall_i \in [1, K]$
\For{$n=1$ to $\mathrm{N}_\mathrm{pilot}$}
    \State \textsc{Pilot symbol: } $\mathbf{s}_{\mathrm{pilot},n} = [s_{n,1}, s_{n,2}, \dots , s_{n,K}]$
    \For{\acp{TX} $i=1$ to $K$}
        \State \textsc{Transmit: } TX$_i$ $\rightarrow$ $s_{n,i}N_{\mathrm{TX},i}$ at $t_{\mathrm{off},i}$
        \State \textsc{Detect: } \ac{RX} uses $\tau_i^\mathbf{\tilde s_{n,i-1}}$ to obtain $\hat{s}_{n,i}$, Eq.~\ref{eq:detection_noma}
        \If{$\hat{s}_{n,i}\neq s_{n,i}$ \textbf{AND} $s_{n,i}=0$}
            \State $\tau_i^\mathbf{\tilde s_{n,i-1}} \gets \tau_i^\mathbf{\tilde s_{n,i-1}} + \Delta \tau$
        \ElsIf{$\hat{s}_{n,i}\neq s_{n,i}$ \textbf{AND} $s_{n,i}=1$}
            \State $\tau_i^\mathbf{\tilde s_{n,i-1}} \gets \tau_i^\mathbf{\tilde s_{n,i-1}} - \Delta \tau$
        \EndIf
    \EndFor
\EndFor
\State \textsc{Output:} $\mathcal{T}_i^\mathrm{NOMA} = \{\tau_i^\mathbf{\tilde s_{i-1}}\}\ \forall_i \in [1, K]$
\end{algorithmic}
\end{algorithm}

To heuristically optimize the detection thresholds, we first assume that the other parameters, namely $t_{\mathrm{off},i}$ and $N_{\mathrm{TX},i}$ are fixed for all TX$_i$.
The sequence of pilot symbol vectors known to the \ac{RX} and all \acp{TX} is denoted as $\mathbf{S}_\mathrm{pilot} = [s_{n,i}\in\{0,1\}\ \mathrm{for}\ 1 \leq n \leq \mathrm{N}_\mathrm{pilot}, 1\leq i \leq K]$. The pilot symbol vector of a single pilot iteration is $\mathbf{s}_{\mathrm{pilot},n} = [s_{n,i}\in\{0,1\}\ \mathrm{for}\ 1\leq i \leq K]$.
The pilot symbol procedure consists of executing the \ac{DBMC-NOMA} scheme as described in Section~\ref{sec:system_model}, i.e. all \acp{TX} send the bit $s_{n,i}$ for pilot symbol $n$  and the bit from each \ac{TX} within the symbol period is decoded by the \ac{RX} in sequence. In addition to the standard procedure, the detection thresholds $\tau_i^\mathbf{\hat s_{i-1}}$ are adjusted after the transmission, sampling, and decoding of each pilot symbol, starting from an initial value $\tau_{i,\mathrm{init}}^\mathbf{\hat s_{i-1}}\ \forall_i$.
First, the detected symbol for TX$_i$ is compared to the correct symbol in the pilot sequence. If the symbol was detected correctly, the threshold stays the same. If the symbol was incorrectly detected as a '1', the threshold must be increased, so the detection of a '0' becomes more likely next time. Mirroring this behavior, the threshold is decreased if the symbol was incorrectly detected as a '0'.

During the pilot symbol process, the thresholds $\tau_i^\mathbf{\tilde{\mathbf{s}}_{n,i-1}}$ are applied for detection, where $\tilde{\mathbf{s}}_{n,i-1} = [s_{n,j}\in\{0,1\}\ \mathrm{for}\ 1\leq j \leq i-1]$ are the pilot symbols up to and including TX$_{i-1}$, as opposed to the detected symbol vector $\hat{\mathbf{s}}_{i-1}$. This ensures that the thresholds are not optimized for the temporary suboptimal situation at the outset, but with the assumption of correct detection for all previous \acp{TX}.
The scheme is described in detail in Algorithm~\ref{alg:threshold}.

\subsection{Avoiding Worst-Case-Offset Scenarios}

{\begin{algorithm}[t]
\caption{Worst-Case-Offset Avoidance Mechanism (WCAM)}
\label{alg:wcam}
\begin{algorithmic}
\footnotesize
\State Obtain pilot decisions $\hat{s}_{n,i}$ for all pilot vectors $\mathbf{s}_{\mathrm{pilot},n}$, $n=1,\dots,N_{\mathrm{pilot}}$, and all $i=1,\dots,K$
\State $\mathcal{P} \gets \{N_{\mathrm{pilot}}-N_{\mathrm{pilot}}^{\mathrm{sel}}+1,\dots,N_{\mathrm{pilot}}\}$
\State $C_{\mathrm{eq}}, C_{\mathrm{mix}}, N_{\mathrm{eq}}, N_{\mathrm{mix}} \gets 0$
\For{$n \in \mathcal{P}$}
    \For{$i = 1$ to $K$}
        \State $e_{n,i} \gets \mathbbold{1}\{\hat{s}_{n,i} \neq s_{n,i}\}$
    \EndFor
    \State \rev{$E_n \gets \sum_{i=1}^{K} e_{n,i}$}
    \If{$\mathbf{s}_{\mathrm{pilot},n} = \mathbf{0}$ \textbf{or} $\mathbf{s}_{\mathrm{pilot},n} = \mathbf{1}$}
        \State $N_{\mathrm{eq}} \gets N_{\mathrm{eq}} + K$
        \State $C_{\mathrm{eq}} \gets C_{\mathrm{eq}} + \rev{E_n}$
    \Else
        \State $N_{\mathrm{mix}} \gets N_{\mathrm{mix}} + K$
        \State $C_{\mathrm{mix}} \gets C_{\mathrm{mix}} + \rev{E_n}$
    \EndIf
\EndFor
\If{$C_{\mathrm{mix}} \ge \rho_{\mathrm{mix}} N_{\mathrm{mix}}$ \textbf{and} $C_{\mathrm{eq}} \le \rho_{\mathrm{eq}} N_{\mathrm{eq}}$ \textbf{and} $C_{\mathrm{mix}}-C_{\mathrm{eq}} \ge \rho_{\Delta} N_{\mathrm{mix}}$}
    \State \textsc{RX: Send WCAM beacon}
    \For{$i = 1$ to $K$}
        \State TX$_i$: $t_{\mathrm{off},i} \gets (t_{\mathrm{off},i} + \Delta_{s,i}) \bmod T$
    \EndFor
\Else
    \State offsets remain unchanged
\EndIf
\end{algorithmic}
\color{black}
\end{algorithm}}

To address the choice of synchronization offsets, we note that the results in Section~\ref{sec:ma_comparison} have shown that it is sufficient to avoid a few narrow worst-case offset regions in order to approach the upper-bound \ac{MDMA} performance. Consequently, our approach is not one of iterative optimization of $t_{\mathrm{off},i}$, but of targeted worst-case avoidance.

The key observation underlying the \ac{WCAM} is that unfavorable offset constellations mainly manifest themselves through excess errors on \emph{mixed} pilot vectors, i.e., pilot vectors for which some \acp{TX} transmit a `1' while others transmit a `0'. For \ac{OOK}, this is the case in which \ac{MAI} is most detrimental, since molecules emitted by active \acp{TX} can impair the detection of inactive ones. In contrast, the all-equal pilot vectors $\mathbf{0}$ and $\mathbf{1}$ provide a reference case: for $\mathbf{0}$, no \ac{MAI} is present, while for $\mathbf{1}$, all \acp{TX} benefit from the increased received signal level. Hence, a likely worst-case offset constellation is characterized not by a high error rate alone, but by a distinctly higher error level on mixed pilots than on all-equal pilots.

Based on this idea, the \ac{WCAM} in Algorithm~\ref{alg:wcam} evaluates the detected pilot block at the \ac{RX}. It counts erroneous \ac{TX}-bit decisions individually, i.e. for each selected pilot vector $\mathbf{s}_{\mathrm{pilot},n}$, the indicator $e_{n,i} = \mathbbold{1}\{\hat{s}_{n,i} \neq s_{n,i}\}$ is formed for every \ac{TX}~$i$. The \ac{RX} then separates the selected pilot vectors into all-equal cases, $\mathbf{s}_{\mathrm{pilot},n} \in \{\mathbf{0},\mathbf{1}\}$, and mixed cases, $\mathbf{s}_{\mathrm{pilot},n} \notin \{\mathbf{0},\mathbf{1}\}$, and accumulates the corresponding error counts $C_{\mathrm{eq}}$ and $C_{\mathrm{mix}}$, as well as the associated denominators $N_{\mathrm{eq}}$ and $N_{\mathrm{mix}}$. $N_{\mathrm{eq}}$ and $N_{\mathrm{mix}}$ denote the corresponding numbers of evaluated \ac{TX}-bit decisions.

To ensure that the trigger decision reflects the current operating state, only the most recent $N_{\mathrm{pilot}}^{\mathrm{sel}}$ pilot vectors are used for the \ac{WCAM} statistic. A trigger is generated only if three guards are satisfied simultaneously: i) the number of mixed-pilot errors is sufficiently large, ii) the number of all-equal-pilot errors remains limited, and iii) the excess of mixed-pilot errors over all-equal-pilot errors exceeds a minimum margin. 
The \ac{WCAM} beacon is sent if
\begin{equation}
    C_{\mathrm{mix}} \ge \rho_{\mathrm{mix}} N_{\mathrm{mix}},\ 
    C_{\mathrm{eq}} \le \rho_{\mathrm{eq}} N_{\mathrm{eq}},\ 
    C_{\mathrm{mix}} - C_{\mathrm{eq}} \ge \rho_{\Delta} N_{\mathrm{mix}}.
\end{equation}

Here, $\rho_{\mathrm{mix}}$, $\rho_{\mathrm{eq}}$, and $\rho_{\Delta}$ are fixed design fractions that determine the required mixed-pilot error level, the tolerated all-equal-pilot error level, and the minimum mixed-versus-equal error margin, respectively. If the \ac{RX} sends the \ac{WCAM} beacon, the \acp{TX} apply a predetermined offset update sequence, which is known to both \acp{TX} and \ac{RX} and can be encoded directly in the pilot sequence $\mathbf{S}_{\mathrm{pilot}}$. The individual additive offset values $\Delta_{s,i}$ are drawn uniformly from $\mathcal{U}[0,\Delta_{\mathrm{s,max}}]$, where $\Delta_{\mathrm{s,max}}$ denotes the \ac{WCAM} delay bound. For $\Delta_{\mathrm{s,max}} = 0$, the \ac{WCAM} is deactivated.

\subsection{Optimizing the Emitted Number of Molecules}

We will now describe the optimization step for the number of emitted molecules $N_{\mathrm{TX},i}$, as proposed in~\cite{wietfeldErrorProbabilityOptimization2024c}, for a system of 2 \acp{TX}. We will limit the consideration to this scenario, since any extension would cause a large increase in description complexity, and we will show later that the mechanism is not necessary when using the \ac{WCAM} above.
Now, $\tau_i^\mathbf{\hat s_{i-1}}$ and $t_{\mathrm{off},i}$ remain fixed and the values of $N_{\mathrm{TX},i}$ are adjusted based on the transmission, sampling and decoding of the pilot sequence.
Given the molecule budget $N_\mathrm{TX,max}$, one of the \acp{TX} should emit exactly the maximum number of molecules, while the other emits a number equal or below. Therefore, for the heuristic, we assume that TX$_1$ is assigned $N_\mathrm{TX,1} = N_\mathrm{TX,max}$, while $N_\mathrm{TX,2}$ is optimized via the scheme, starting from an initial value $N_\mathrm{TX,init}$.
In~\cite{wietfeldErrorProbabilityOptimization2024c}, we showed that this works as long as the distances are sufficiently similar and if the \ac{RX} can communicate the change in $N_\mathrm{TX,2}$ to the correct \ac{TX}.
Then, there is a set of decision rules governing the adjustment in $N_\mathrm{TX,2}$ after determining the detected and correct bits $\hat{s}_{n,1}, \hat{s}_{n,2}$ and $s_{n,1}, s_{n,2}$, respectively. 

If $s_{n,2}=0$, $N_\mathrm{TX,2}$ does not influence the result and there is no reason for adjustment. For $s_{n,2}=1$, we will look at two example cases for the adjustment rules. The entire scheme is described in Algorithm~\ref{alg:numMol}.
If $\hat{s}_{n,1} \neq s_{n,1} = 0$ (incorrectly detected a '1'), and $\hat{s}_{n,2} = s_{n,2} = 1$ (correctly detected a '1'), we can infer that we observed enough molecules at the \ac{RX} to correctly detect a '1' for TX$_2$, but the \ac{MAI} for TX$_1$ seems to have caused too many molecules to arrive, so that we misclassified its symbol. Therefore, $N_\mathrm{TX,2}$ should be reduced.
If $\hat{s}_{n,1} = s_{n,1} = 0$ (correctly detected a '0'), and $\hat{s}_{n,2} \neq s_{n,2} = 1$ (incorrectly detected a '0'), we can infer that the \ac{MAI} for the TX$_1$ detection was not too high, but the received molecules for TX$_2$ were not high enough to cross the threshold for a '1'. Therefore, $N_\mathrm{TX,2}$ should be increased.

\begin{algorithm}[t]
\caption{Number of Molecules Optimization Algorithm}\label{alg:numMol}
\begin{algorithmic}
\footnotesize

\State \textsc{Input:} $N_{\mathrm{TX},2}$
\For{$n=1$ to $\mathrm{N}_\mathrm{pilot}$}
    \State \textsc{Pilot symbol: } $\mathbf{s}_{\mathrm{pilot},n} = [s_{n,1}, s_{n,2}, \dots , s_{n,K}]$
    \For{\acp{TX} $i=1$ to $K$}
        \State \textsc{Transmit: } TX$_i$ $\rightarrow$ $s_{n,i}N_{\mathrm{TX},i}$ at $t_{\mathrm{off},i}$
        \State \textsc{Detect: } \ac{RX} uses $\tau_i^\mathbf{\tilde s_{n,i-1}}$ to obtain $\hat{s}_{n,i}$, Eq.~\ref{eq:detection_noma}
        \If{$s_{n,2} = 1$}
            \If{$s_{n,1}=0$ \textbf{AND} $\hat{s}_{n,1}\neq s_{n,1}$ \textbf{AND} $\hat{s}_{n,2} = s_{n,2}$}
                \State $N_{\mathrm{TX},2} \gets N_{\mathrm{TX},2}\cdot(1-\alpha_\mathrm{N})\ \mathrm{with\ probability\ } 1-p_\mathrm{e,f}$
            \EndIf
            \If{[$s_{n,1}=0$ \textbf{AND} $\hat{s}_{n,1}= s_{n,1}$ \textbf{AND} $\hat{s}_{n,2} \neq s_{n,2}$] \\$\phantom{a}$\textbf{OR} [$s_{n,1}=1$ \textbf{AND} $\hat{s}_{n,1}\neq s_{n,1}$ \textbf{AND} $\hat{s}_{n,2} \neq s_{n,2}$] \\$\phantom{a}$\textbf{OR} [$s_{n,1}=1$ \textbf{AND} $\hat{s}_{n,1}= s_{n,1}$ \textbf{AND} $\hat{s}_{n,2} \neq s_{n,2}$] }
                \State $N_{\mathrm{TX},2} \gets N_{\mathrm{TX},2}\cdot(1+\alpha_\mathrm{N})\ \mathrm{with\ probability\ } 1-p_\mathrm{e,f}$
            \EndIf
        \EndIf
    \EndFor
\EndFor
\State \textsc{Output:} $N_{\mathrm{TX},2}$ 
\end{algorithmic}
\end{algorithm}

\emph{Feedback Mechanism Model: }
To incorporate the effects of a feedback channel, we propose a simple model. Assuming the \ac{RX} communicates the necessary adjustment back to TX$_2$ via a separate control channel molecule and encodes the information using orthogonal binary sequences decoded via correlation, a binary erasure channel appears as an appropriate simplification. Due to the lack of interference from the main communication molecule channel, the information is either recovered correctly, or erased entirely, when the correlation is unsuccessful.
To adjust $N_\mathrm{TX,2}$, a multiplicative model is used, such that
    \begin{equation}
        N_\mathrm{TX,2} \leftarrow \begin{cases}
            N_\mathrm{TX,2} &\text{with prob. } p_\mathrm{e,f}\\
            N_\mathrm{TX,2}\cdot(1\pm \alpha_\mathrm{N})\quad &\text{else},
        \end{cases}
    \end{equation}
with the \emph{number of molecules multiplier} $\alpha_\mathrm{N}$.
If not otherwise specified, $p_\mathrm{e,f} = 0$.

\subsection{DBMC-aNOMAly Protocol}
Neither Algorithm~\ref{alg:threshold},~\ref{alg:wcam}, nor~\ref{alg:numMol} can itself find a jointly favorable operating point. Therefore, we propose a joint optimization scheme that combines the individual mechanisms across multiple iterations while keeping the adaptation in each step focused on a limited set of parameters. We define a number of joint algorithm iterations $N_{\mathrm{iter}}$. In every iteration, the detection-threshold optimization in Algorithm~\ref{alg:threshold} is carried out using $N_{\mathrm{pilot}}$ pilot symbols. In parallel, the \ac{WCAM} in Algorithm~\ref{alg:wcam} is evaluated on the same pilot block and, if triggered, updates the synchronization offsets $t_{\mathrm{off},i}$ via the beacon-based offset shift.

If the optimization of the emitted number of molecules is included, Algorithm~\ref{alg:numMol} is not applied simultaneously with Algorithm~\ref{alg:threshold} in the same adaptation step. Instead, Algorithm~\ref{alg:threshold} and Algorithm~\ref{alg:numMol} are applied alternatingly across successive joint iterations, such that only one of the two parameter-update mechanisms acts at a time, while Algorithm~\ref{alg:wcam} continues to run in parallel throughout. Hence, the standard version of \ac{DBMC-aNOMAly} consists of Algorithm~\ref{alg:threshold} together with Algorithm~\ref{alg:wcam}, whereas the extended version additionally alternates with Algorithm~\ref{alg:numMol} when the optimization of $N_{\mathrm{TX},i}$ is enabled.

\section{Protocol Evaluation}\label{sec:protocol_eval}

\begin{table}[t]
\caption{Parameters for the Protocol Evaluation}
\label{tab:protocol_parameters}
\resizebox{\columnwidth}{!}{%
\begin{tabular}{@{}lll@{}}
\toprule
\textbf{Parameter} & \textbf{Symbol} & \textbf{Values ($\underline{\mathrm{Default}}$)} \\ 
\midrule

\multicolumn{3}{@{}l}{\textbf{Communication System}}\\
\addlinespace[2pt]
Number of TXs            & $K$                  & $2, 3, \underline{4}, 5$ \\
TX distances              & $d_i$                & $\{8, \underline{10}, 12\}\,\qty{}{\micro\meter}$ \\
RX radius                 & $r$                  & $\qty{1}{\micro\meter}$ \\
Diffusion coefficient     & $D$                  & $\qty{e-9}{\meter\squared\per\second}$ \\
Symbol period             & $T$                  & $1\,\unit{\second}$ \\
\ac{ISI} symbols          & $L$                  & $1$ \\
Signaling-molecule-to-noise ratio & SNR          & $\{-10,0,10,\underline{\infty}\}\,\unit{\decibel}$ \\
Sampling jitter           & $\Delta_\mathrm{p}$  & $\{\underline{0}, 0.05, 0.1\}\,T$ \\
Molecule budget per \acs{TX} & $N_\mathrm{TX,max}$ & $\qty{e6}{\mathrm{molecules}}$ \\
Feedback erasure probability & $p_\mathrm{e,f}$ & $\{\underline{0},0.5,0.99\}$ \\

\addlinespace[4pt]
\multicolumn{3}{@{}l}{\textbf{Protocol Parameters}}\\
\addlinespace[2pt]
Number of seeds           & $\mathrm{N}_\mathrm{seed}$ & $\{\underline{100}, 500, 1000\}$ \\
Number of pilot symbols   & $\mathrm{N}_\mathrm{pilot}$ & $100$ \\
Number of iterations      & $\mathrm{N}_\mathrm{iter}$  & $1000$ \\
Threshold step            & $\Delta \tau$             & $\qty{1}{\mathrm{molecule}}$ \\
Number of molecules multiplier & $\alpha_\mathrm{N}$  & $0.1$ \\
Initial thresholds        & $\tau_{j,\mathrm{init}}^{\mathbf{\hat{s}}_{j-1}}$ & $ 1\,\forall_j$  \\
Initial number of molecules & $N_{\mathrm{TX,init}}$ & $10^6\,\qty{}{\mathrm{molecules}}$ \\

\addlinespace[4pt]
\multicolumn{3}{@{}l}{\textbf{WCAM Parameters}}\\
\addlinespace[2pt]
Delay bound     & $\Delta_\mathrm{s,max}$ & $\{0, 0.1, 0.5, \underline{1}\}\,\unit{\second}$ \\
Recent pilots selected & $N_{\mathrm{pilot}}^{\mathrm{sel}}$ & $N_{\mathrm{pilot}}/2$\\
Mixed-error fraction threshold & $\rho_{\mathrm{mix}}$ & 0.1 \\
All-equal-error fraction threshold & $\rho_{\mathrm{eq}}$ & 0.25 \\
Mixed-vs-equal fraction threshold & $\rho_{\Delta}$ & 0.1 \\
\bottomrule
\end{tabular}%
}
\end{table}

This section evaluates \ac{DBMC-aNOMAly} by simulation using randomly generated pilot sequences and the Poisson channel model from Section~\ref{sec:system_model}, including the sampling-jitter model from Subsection~\ref{subsec:communication_system}. We vary the network size $K$, noise level, sampling jitter, and inclusion of Algorithm~\ref{alg:numMol}. We also evaluate pilot overhead, protocol efficiency, and changing parameters during runtime.

For each result, the full protocol run has been repeated at least 100 times, while for high-variance scenarios, i.e. for low SNR or large sampling jitter, we utilized up to 1000 repetitions.
We first analyze the \ac{WCAM} trigger probability, specificity and the impact on \ac{BEP} trajectories. Then, results for the systematic parameter analysis will be presented.

An overview of the simulation parameters can be found in Table~\ref{tab:protocol_parameters}. The default values, indicated by the underline, are utilized unless otherwise stated. As discussed in Section~\ref{sec:system_model}, the values of distances $d_i$ and \ac{RX} radius $r$ allow the application of the \ac{UCA}, since the condition $r < 0.15 d_i$ is valid for all considered scenarios.

\subsection{Protocol Evaluation Metrics}

To evaluate the DBMC-aNOMAly protocol, we use three complementary groups of metrics. First, the main convergence metric is the system \ac{BEP} $P_{\mathrm{e,sys}}$, as defined in Section~\ref{sec:bep} and used throughout the analytical evaluation in Section~\ref{sec:ma_comparison}. In the following parameter sweeps, $P_{\mathrm{e,sys}}$ therefore quantifies how reliably and how quickly the protocol improves the communication performance over the optimization iterations.

Second, to characterize the behavior of the \ac{WCAM} itself, we use mechanism-specific statistics. 
% Together, these metrics indicate whether the \ac{WCAM} activates selectively in potentially unfavorable offset constellations and whether its interventions are typically followed by an immediate improvement.
The trigger probability in iteration $m$ denotes the fraction of protocol runs in which the \ac{WCAM} beacon is sent in iteration $m$. The minimum pairwise offset difference is defined as
\begin{multline}
    \delta_{\min}(m) = \min_{i<j} \min\!\left( \left| t_{\mathrm{off},i}(m)-t_{\mathrm{off},j}(m)\right|,\right.\\ \left.T-\left| t_{\mathrm{off},i}(m)-t_{\mathrm{off},j}(m)\right| \right),
\end{multline}
i.e., the smallest distance between any two \acp{TX} in a given iteration. Small values of $\delta_{\min}$ indicate that at least one \ac{TX} pair is close to a potentially unfavorable offset constellation. For each trigger event, we evaluate the immediate post-trigger \ac{BEP} reduction
\begin{equation}
    \Delta P_{\mathrm{e,sys}}^\mathrm{trig}
    =
    P_{\mathrm{e,sys}}^{\mathrm{pre}} - P_{\mathrm{e,sys}}^{\mathrm{post}},
\end{equation}
where $P_{\mathrm{e,sys}}^{\mathrm{pre}}$ and $P_{\mathrm{e,sys}}^{\mathrm{post}}$ denote the system \ac{BEP} directly before and after the trigger, respectively.

Third, to assess the practical efficiency of the protocol for finite payloads, we use end-to-end metrics that explicitly account for the pilot overhead. For this purpose, we reuse the \ac{MI}-based throughput proxy $\mathcal{I}_\mathrm{sys}$ from Section~\ref{sec:ma_comparison}. Let $m_\mathrm{hit}$ denote the first protocol iteration for which a target \ac{BEP} $P_\mathrm{e,sys}^\mathrm{target}$ is reached, and let $N_{\mathrm{pilot,used}}=m_\mathrm{hit}N_\mathrm{pilot}$ be the corresponding number of pilot symbols spent during adaptation. For a subsequent payload of $N_\mathrm{pay}$ symbols, we define the net throughput as
\begin{equation}
    \mathcal{I}_{\mathrm{sys,net}}(N_\mathrm{pay})
    =
    \mathcal{I}_{\mathrm{sys}}(m_\mathrm{hit})
    \frac{N_\mathrm{pay}}{N_\mathrm{pay}+N_{\mathrm{pilot,used}}},
\end{equation}
where $\mathcal{I}_{\mathrm{sys}}(m_\mathrm{hit})$ denotes the \ac{MI}-based throughput at the first target-reaching iteration. In addition, the pilot time to target is given by
\begin{equation}
    T_\mathrm{hit}=N_{\mathrm{pilot,used}}T.
\end{equation}
Hence, $\mathcal{I}_{\mathrm{sys,net}}$ captures the trade-off between improved post-adaptation communication performance and the pilot overhead required to obtain it, while $T_\mathrm{hit}$ quantifies the corresponding adaptation latency.

\subsection{Investigating the WCAM}

\begin{figure}[t]
    \begin{subfigure}{\linewidth}
    \centering
        \includegraphics[width=\linewidth]{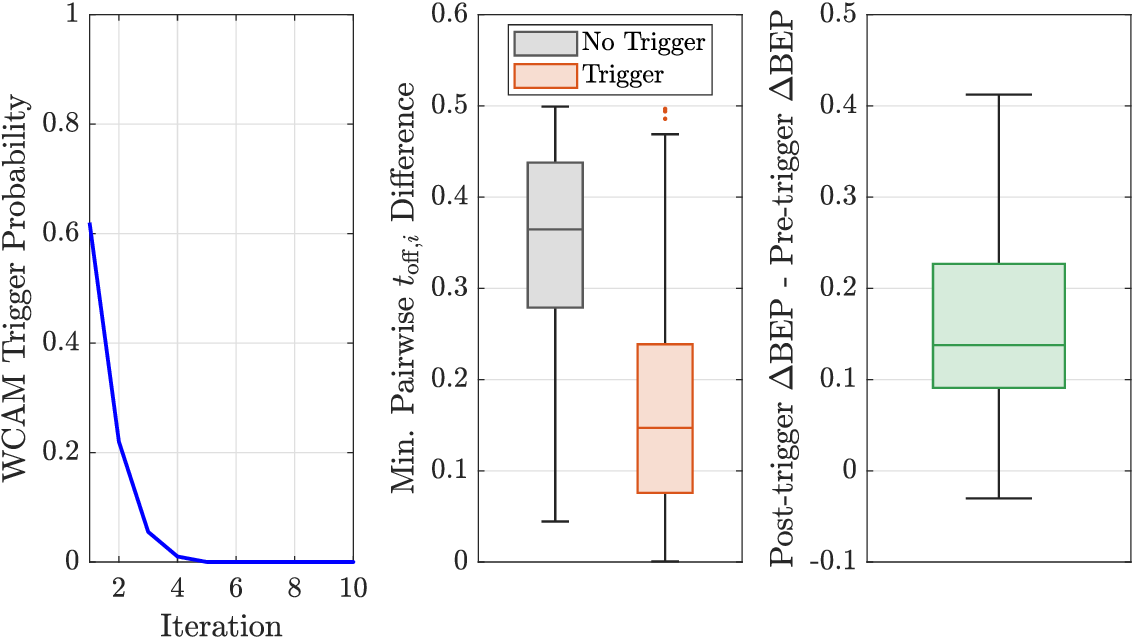}
        \caption{$K=2$, SNR = $\infty$ dB}
        \label{fig:wcam_stats:subfig:2tx}
    \end{subfigure}
    \hfill
    \begin{subfigure}{\linewidth}
    \centering
        \includegraphics[width=\linewidth]{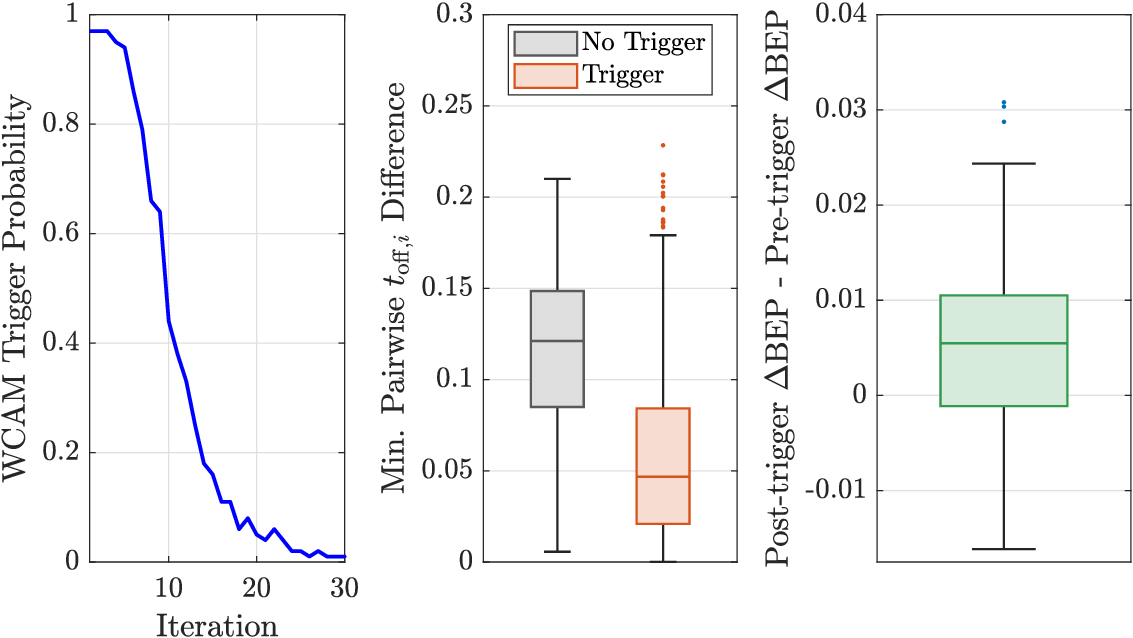}
        \caption{$K=4$, SNR = $\infty$ dB}
        \label{fig:wcam_stats:subfig:4tx}
    \end{subfigure}
    \hfill
    \begin{subfigure}{\linewidth}
        \includegraphics[width=0.955\linewidth]{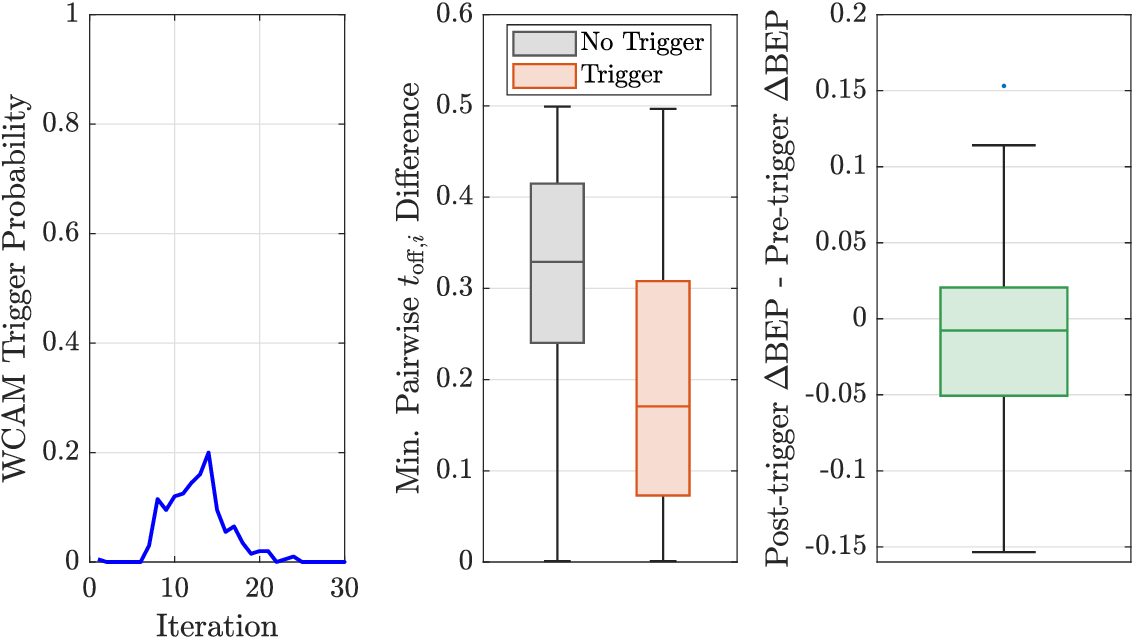}
        \caption{$K=2$, SNR = 0 dB}
        \label{fig:wcam_stats:subfig:SNR0}
    \end{subfigure}
    \caption{Statistical characterization of the \ac{WCAM}. Left: probability that the \ac{WCAM} beacon is triggered in a given protocol iteration. Middle: distribution of the minimum pairwise cyclic offset difference $\delta_{\min}$, conditioned on whether a trigger occurred in that iteration. Right: distribution of the immediate post-trigger \ac{BEP} reduction $\Delta P_{\mathrm{e,sys}}^\mathrm{trig}=P_{\mathrm{e,sys}}^{\mathrm{pre}}-P_{\mathrm{e,sys}}^{\mathrm{post}}$. All unspecified parameters follow the default values in Table~\ref{tab:protocol_parameters}.}
    \label{fig:wcam_stats}
\end{figure}

First, we investigate whether the \ac{WCAM} has the desired effect of avoiding worst-case offset scenarios. Since the worst-case offset regions identified in Section~\ref{sec:ma_comparison} do not form a binary class with a sharp boundary, we do not define a strict false-trigger or miss-trigger rate. Instead, we characterize the \ac{WCAM} statistically by three quantities shown in Fig.~\ref{fig:wcam_stats}: the trigger probability, the distribution of minimum pairwise offset difference $\delta_\mathrm{min}$, and the immediate \ac{BEP} reduction after a trigger $\Delta P_{\mathrm{e,sys}}^\mathrm{trig}$.

Fig.~\ref{fig:wcam_stats:subfig:2tx} shows $K=2$ and $\mathrm{SNR}=\infty\, \unit{\decibel}$. The trigger probability is concentrated in the first few iterations and quickly approaches zero, so the \ac{WCAM} mainly acts during the initial phase without perturbing settled constellations. The trigger-conditioned distribution of $\delta_{\min}$ is shifted towards smaller values than the no-trigger case, indicating selective activation in more critical offset constellations. The mostly positive post-trigger \ac{BEP} reduction shows that interventions are typically followed by an immediate gain.

The same qualitative behavior remains visible for $K=4$ in Fig.~\ref{fig:wcam_stats:subfig:4tx}. Here, the trigger probability starts higher and decays more gradually, which is expected since a larger network creates more pairwise offset constellations and more opportunities for at least one unfavorable pair. Nevertheless, triggered iterations again exhibit systematically smaller values of $\delta_{\min}$ than non-triggered ones, indicating that the \ac{WCAM} remains selective also beyond the $K=2$ case. The post-trigger \ac{BEP} reduction is smaller on average than in the $K=2$ case, but still predominantly positive.

For the noisy case with $K=2$ and $\mathrm{SNR}=0\,\unit{\decibel}$ in Fig.~\ref{fig:wcam_stats:subfig:SNR0}, the trigger probability becomes weaker and less regular, and the separation between the trigger and no-trigger distributions of $\delta_{\min}$ is reduced. This indicates that noise degrades the identifiability of unfavorable offset constellations, which is consistent with the reduced \ac{WCAM} stability observed later in the full protocol results. Correspondingly, the distribution of the post-trigger \ac{BEP} reduction broadens and includes more negative values, showing that under noisy conditions the immediate effect of a trigger becomes less reliable. However, $\Delta P_{\mathrm{e,sys}}^\mathrm{trig}$ does not consider the long-term effects of a \ac{WCAM} intervention. We will see later in this section that it remains advantageous even for scenarios with significant noise.

Overall, Fig.~\ref{fig:wcam_stats} shows that the \ac{WCAM} is effective beyond selected examples: under favorable conditions, it triggers during early high-risk constellations, is associated with small minimum pairwise offset differences, and is typically followed by an immediate \ac{BEP} reduction. The selectivity remains visible for larger $K$ and degrades predictably for lower \ac{SNR}.

\begin{figure}[t]
    \centering
    \includegraphics[width=0.8\linewidth]{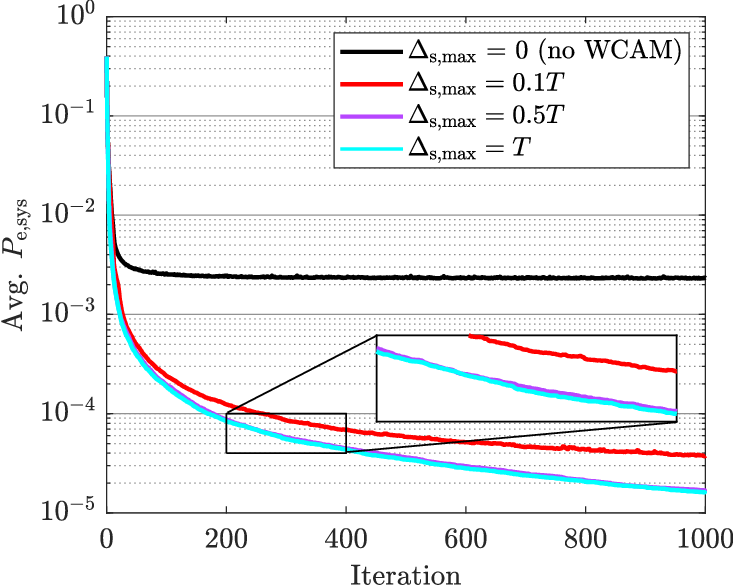}
    \caption{Average \ac{BEP}, $P_\mathrm{e,sys}$, across 1000 iterations of the \ac{DBMC-aNOMAly} protocol. Results shown for different values of the \ac{WCAM} delay bound $\Delta_\mathrm{s,max}$, determining the maximum offset induced by the \ac{WCAM} beacon. $K=2$. All other parameters according to the default in Table~\ref{tab:protocol_parameters}.}
    \label{fig:delayBound}
\end{figure}

In Fig.~\ref{fig:delayBound}, the resulting average \ac{BEP} trajectories over 1000 iterations are shown for different values of the \ac{WCAM} delay bound $\Delta_\mathrm{s,max}$ between 0 (no \ac{WCAM}) and $T$.
We can see a contrast between the scenarios with and without \ac{WCAM}, as for $\Delta_\mathrm{s,max} = 0$, the large number of badly performing runs causes $P_\mathrm{e,sys}$ to approach a constant value very early, while for $\Delta_\mathrm{s,max}>0$, the optimization keeps reducing $P_\mathrm{e,sys}$ much further by up to two orders of magnitude.
The zoomed-in section highlights the difference between different values of $\Delta_\mathrm{s,max}$, i.e. different magnitudes of the offset adjustment after the \ac{WCAM} beacon is sent out. We can observe that the differences are small, but the performance improves with larger values of $\Delta_\mathrm{s,max}$.
Therefore, we will choose $\Delta_\mathrm{s,max} = T$ as the default going forward.

\subsection{Parameter sweeps and convergence behavior}

After establishing the qualitative behavior and selectivity of the \ac{WCAM}, we now turn to the overall convergence behavior of the DBMC-aNOMAly protocol under systematic parameter variations. We keep the protocol structure fixed and evaluate how key system and protocol parameters affect the \ac{BEP} trajectory over the optimization iterations. Unless otherwise stated, we compare the standard protocol without \ac{WCAM}, i.e. $\Delta_\mathrm{s,max}=0$, to the default DBMC-aNOMAly configuration with $\Delta_\mathrm{s,max}=T$, which was identified above as the preferred operating point.

\subsubsection{Varying Network Size}

\begin{figure}[t]
    \centering
    \includegraphics[width=0.8\linewidth]{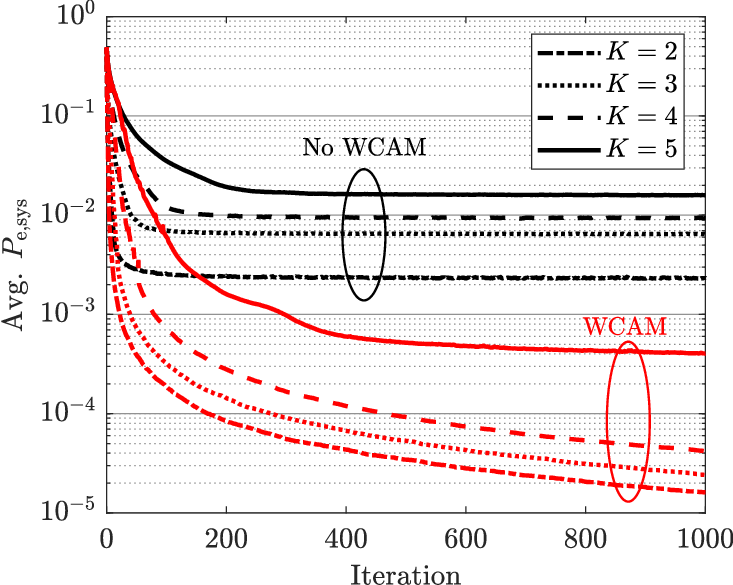}
    \caption{Average \ac{BEP}, $P_\mathrm{e,sys}$, across 1000 iterations of the \ac{DBMC-aNOMAly} protocol. Results shown for different values of the number of \acp{TX}, $K$. No \ac{WCAM} $\rightarrow$ $\Delta_\mathrm{s,max} = 0$. \ac{WCAM}~$\rightarrow$ $\Delta_\mathrm{s,max} = T$. All other parameters according to the default in Table~\ref{tab:protocol_parameters}.}
    \label{fig:numTx}
\end{figure}

In Fig.~\ref{fig:numTx}, the results with and without \ac{WCAM} are presented for network sizes between $K=2$ and $K=5$. The \ac{BEP} trajectories illustrate that while $P_\mathrm{e,sys}$ expectedly decreases for larger networks, the rate of improvement remains roughly similar.
In general, the protocol deals well with increasing $K$, improving upon previous work, which was limited to networks with $K=2$~\cite{wietfeldErrorProbabilityOptimization2024c}.

\subsubsection{Varying Noise Level}

\begin{figure}[t]
    \centering
    \includegraphics[width=0.8\linewidth]{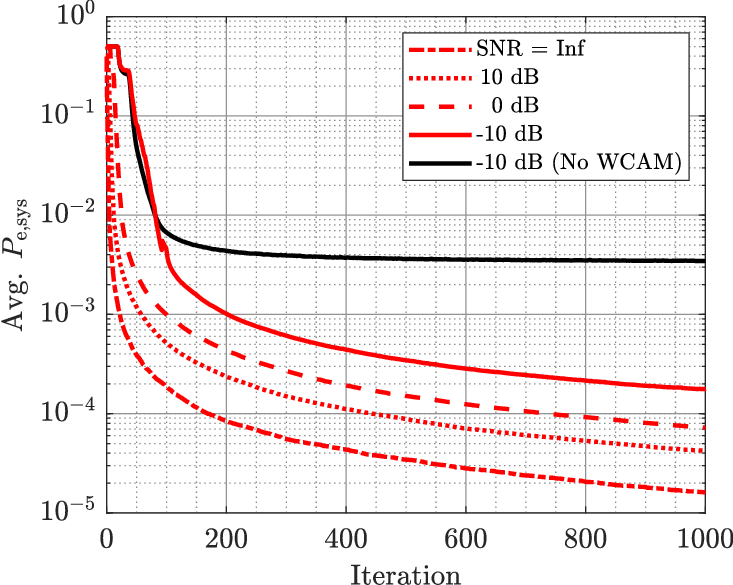}
    \caption{Average \ac{BEP}, $P_\mathrm{e,sys}$, across 1000 iterations of the \ac{DBMC-aNOMAly} protocol. Results shown for different values of the \ac{SNR}. No \ac{WCAM} $\rightarrow$ $\Delta_\mathrm{s,max} = 0$. \ac{WCAM} $\rightarrow$ $\Delta_\mathrm{s,max} = T$. $K=2$. All other parameters according to the default in Table~\ref{tab:protocol_parameters}.}
    \label{fig:snr}
\end{figure}

Next, we evaluate the protocol's ability to deal with higher levels of noise. Fig.~\ref{fig:snr} showcases average $P_\mathrm{e,sys}$ trajectories for \ac{DBMC-aNOMAly} and SNR values between -10 and 10 dB in addition to the case with no noise (infinite SNR). To avoid cluttering of the figure, we only plot the results without the \ac{WCAM} for the case of -10 dB.
The graph highlights that the protocol remains effective under the influence of added noise, and the \ac{WCAM} remains successful for cases down to 0 dB, outperforming the cases without \ac{WCAM} (not shown explicitly). Even for significant noise, i.e. $\mathrm{SNR = -10}$ dB, we observe that the full protocol with \ac{WCAM} outperforms the \textit{No \ac{WCAM}} variant. So, despite higher instability, as shown in Fig.~\ref{fig:wcam_stats:subfig:SNR0}, the \ac{WCAM} remains beneficial.

\subsubsection{Effect of Sampling Jitter}

\begin{figure}[t]
    \centering
    \includegraphics[width=0.8\linewidth]{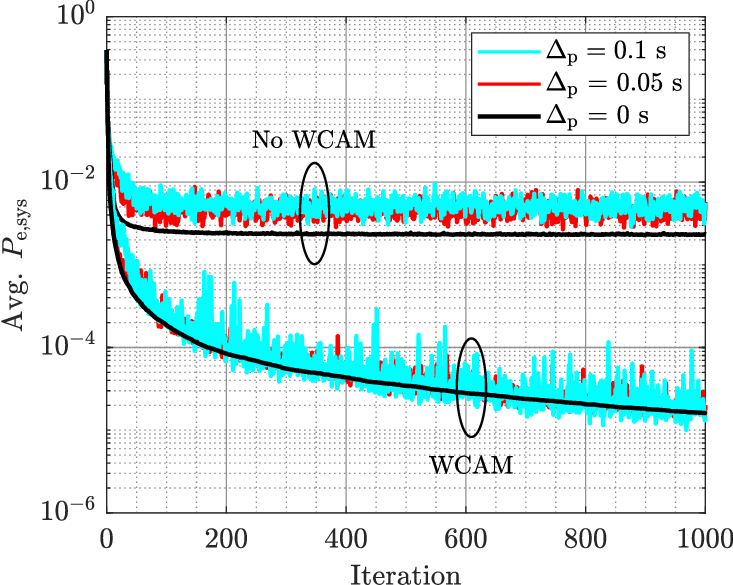}
    \caption{Average \ac{BEP}, $P_\mathrm{e,sys}$, across 1000 iterations of the \ac{DBMC-aNOMAly} protocol. Results shown for different values of the sampling jitter, $\Delta_\mathrm{p}$. No \ac{WCAM} $\rightarrow$ $\Delta_\mathrm{s,max} = 0$. \ac{WCAM} $\rightarrow$ $\Delta_\mathrm{s,max} = T$. $K=2$. All other parameters according to the default in Table~\ref{tab:protocol_parameters}.}
    \label{fig:jitter}
\end{figure}

We have the option to introduce sampling jitter according to the model specified in Section~\ref{subsec:communication_system}. The results for a system with $K=2$ \acp{TX} are shown in Fig.~\ref{fig:jitter}. It depicts the average $P_\mathrm{e,sys}$ trajectory for values of the sampling jitter range $\Delta_\mathrm{p}$ between 0 and $0.1T$.

Comparing the results with and without \ac{WCAM}, we can observe that the \ac{DBMC-aNOMAly} protocol is not affected by the sampling jitter and achieves the same \ac{BEP} improvement. However, more variance is introduced as the performance changes more drastically from iteration to iteration since a misaligned sampling point could lead to temporary inaccurate adjustment. Without the \ac{WCAM} mechanism, sampling jitter leads to a decrease in performance in addition to significant variations.

\subsubsection{Effect of Optimizing the Number of Emitted Molecules}

\begin{figure}[t]
    \centering
    \includegraphics[width=0.8\linewidth]{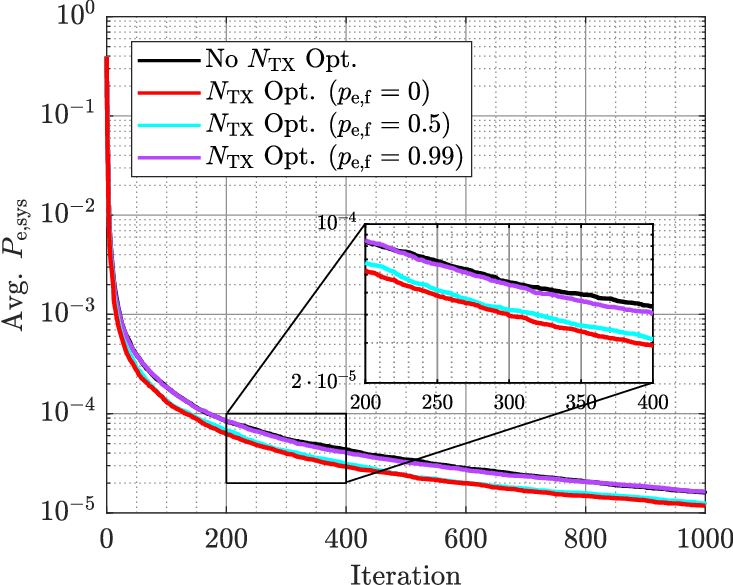}
    \caption{Average \ac{BEP}, $P_\mathrm{e,sys}$, across 1000 iterations of the \ac{DBMC-aNOMAly} protocol. Results shown with and without the $N_\mathrm{TX}$ optimization in Algorithm~\ref{alg:numMol} and for different values of the feedback channel error probability $p_\mathrm{e,f}$. All other parameters according to the default in Table~\ref{tab:protocol_parameters}.}
    \label{fig:numMolOpt}
\end{figure}

In Fig.~\ref{fig:numMolOpt}, we investigate the impact of additional optimization of the number of emitted molecules $N_{\mathrm{TX},i}$, as previously proposed in~\cite{wietfeldErrorProbabilityOptimization2024c} and defined in Algorithm~\ref{alg:numMol}. As discussed in Section~\ref{sec:protocol}, this version is only defined for $K=2$ \acp{TX}. We show the results for the \ac{DBMC-aNOMAly} scheme with and without optimizing $N_{\mathrm{TX},i}$, and apply different values for the feedback error probability $p_\mathrm{e,f}$.

The results show a slight improvement in the performance when Algorithm~\ref{alg:numMol} is included, and the impact of up to $p_\mathrm{e,f} = 0.5$ is very small. For a very unreliable feedback channel ($p_\mathrm{e,f} = 0.99$), performance deteriorates back to the scenario without Algorithm~\ref{alg:numMol}.
Overall, the improvement in $P_\mathrm{e,sys}$ is relatively minor, especially compared to the improvement between the algorithm with and without \ac{WCAM}, as shown in Fig.~\ref{fig:delayBound}. The $N_{\mathrm{TX},i}$ feedback mechanism introduces significant complexity. Therefore, we conclude that in the considered scenario, the performance benefits do not justify the added complexity and effort compared to the standard version of \ac{DBMC-aNOMAly} including Algorithms~\ref{alg:threshold} and~\ref{alg:wcam}.

\subsection{End-to-End Efficiency and Overhead Analysis}

\begin{figure}[t]
    \begin{subfigure}{\linewidth}
    \centering
        \includegraphics[height=3.7cm]{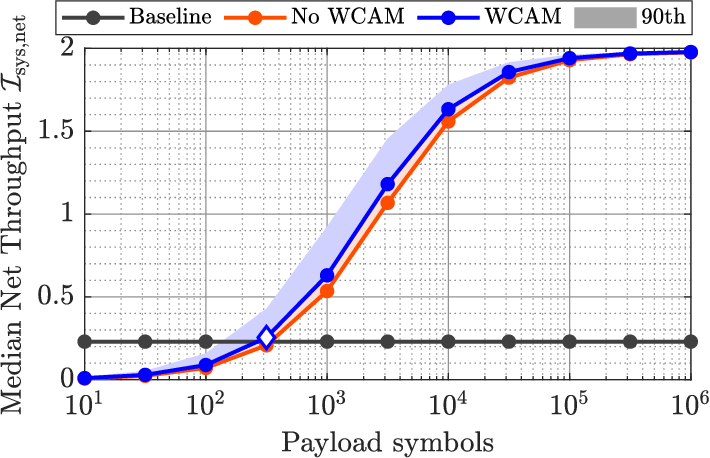}\includegraphics[height=3.7cm]{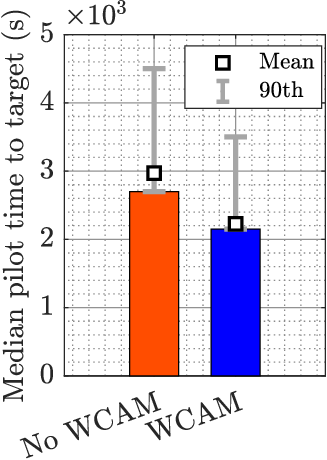}
        \caption{$K=2$, SNR = $\infty$ dB}
        \label{fig:e2e:subfig:2tx}
    \end{subfigure}
    \hfill
    \begin{subfigure}{\linewidth}
    \centering
        \includegraphics[height=3.7cm]{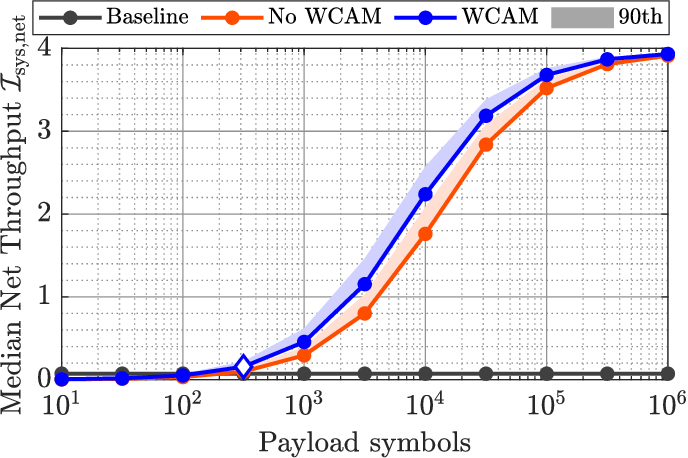}\includegraphics[height=3.7cm]{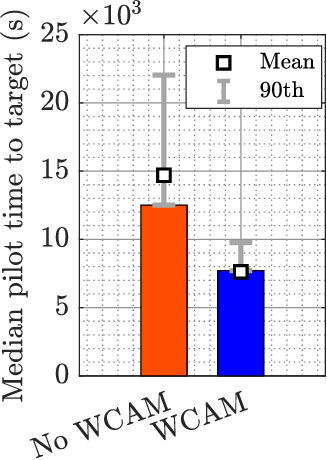}
        \caption{$K=4$, SNR = $\infty$ dB}
        \label{fig:e2e:subfig:4tx}
    \end{subfigure}
    \hfill
    \begin{subfigure}{\linewidth}
    \centering
        \includegraphics[height=3.7cm]{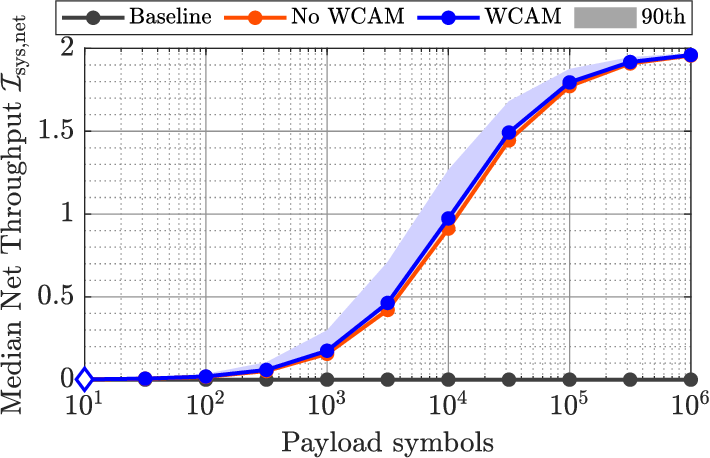}\includegraphics[height=3.7cm]{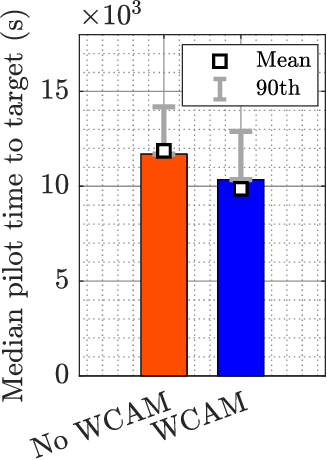}
        \caption{$K=2$, SNR = 0 dB}
        \label{fig:e2e:subfig:SNR0}
    \end{subfigure}
    \caption{End-to-end efficiency of \ac{DBMC-aNOMAly} for three representative protocol scenarios. The left column shows median net \ac{MI}-based throughput $\mathcal{I}_{\mathrm{sys,net}}$ versus payload length after amortizing pilot overhead. The black curve is the iteration-0 baseline; orange and blue indicate threshold adaptation without and with \ac{WCAM}. Shading spans the median to 90th percentile across seeds, and diamonds mark the first plotted payload length with positive median net gain. The right column shows pilot overhead to reach $P_\mathrm{e,sys}^\mathrm{target}=10^{-3}$, with median bars, mean squares, and 90th-percentile whiskers. All other parameters follow Table~\ref{tab:protocol_parameters}.}
    \label{fig:e2e}
\end{figure}

We next evaluate end-to-end efficiency. The previous results quantify convergence of $P_\mathrm{e,sys}$, but not the pilot symbols, time, and molecules spent during adaptation.

Fig.~\ref{fig:e2e} shows this end-to-end view for three representative scenarios. In the left column, the baseline corresponds to the iteration-0 operating point before adaptation, while the two adaptive variants correspond to threshold adaptation without \ac{WCAM} and the full DBMC-aNOMAly protocol with \ac{WCAM}. The right column reports the pilot time required to first reach $P_\mathrm{e,sys}^\mathrm{target} = 10^{-3}$. The figure relates the previously observed \ac{BEP} convergence to practical communication efficiency.

We start with the case $K=2$ and $\mathrm{SNR}=\infty\,\unit{\decibel}$ in Fig.~\ref{fig:e2e:subfig:2tx}. Here, the baseline throughput remains at a comparatively low constant level, whereas both adaptive variants approach a much higher net throughput as the payload length increases. At the same time, for very short payloads the pilot overhead dominates. The break-even point is reached once the payload is sufficiently long to amortize the adaptation phase, with the \ac{WCAM}-based version reaching it earlier. The right-hand plot shows that the \ac{WCAM} reduces the pilot time required to reach the target and, thus, yields a visible end-to-end advantage.

The benefit becomes more pronounced for the larger network with $K=4$ and $\mathrm{SNR}=\infty\,\unit{\decibel}$ shown in Fig.~\ref{fig:e2e:subfig:4tx}. Again, both adaptive variants strongly outperform the baseline once the payload is sufficiently long. However, the gap between \emph{No WCAM} and \emph{WCAM} is now larger over the short- and medium-payload regime. This is consistent with the earlier observation that unfavorable offset constellations become more likely as $K$ increases, making the \ac{WCAM} more valuable. Accordingly, the reduction in pilot time to target is also much more pronounced than in the $K=2$ case.

Finally, Fig.~\ref{fig:e2e:subfig:SNR0} considers the noisy case with $K=2$ and $\mathrm{SNR}=0\,\unit{\decibel}$. Compared to the noiseless case, the pilot overhead required to reach the target increases substantially for both adaptive variants, reflecting the slower and less reliable convergence already observed in Fig.~\ref{fig:snr}. Nevertheless, both protocols provide an immediate net-throughput improvement over the baseline. The gap between the \ac{WCAM}-assisted and non-\ac{WCAM} version becomes smaller than in the noiseless scenarios, which is in line with the statistical characterization in Fig.~\ref{fig:wcam_stats}, where the trigger selectivity of the \ac{WCAM} degraded under noise. Thus, the \ac{WCAM} remains beneficial, but less decisively so than under high-\ac{SNR} conditions.

Overall, the \ac{WCAM} improves end-to-end efficiency primarily by reducing the time required to reach a useful operating point. Since the expected number of emitted molecules is constant for our setup (\ac{OOK}, equiprobable symbols), the reported pilot-symbol time overhead reduction is proportional to the chemical signaling overhead and expected molecule cost reduction. This is relevant for finite payloads, where the pilot overhead is not negligible and becomes more important as the network size increases. The overhead benefits decrease under noisy conditions. The end-to-end analysis confirms that the \ac{BEP} reductions observed above translate into tangible communication-efficiency gains once the finite adaptation cost is taken into account.

\subsection{Changing Parameters during Runtime}

\begin{figure}[t]
    \centering
    \begin{subfigure}{\linewidth}
    \centering 
        \includegraphics[width=0.8\linewidth]{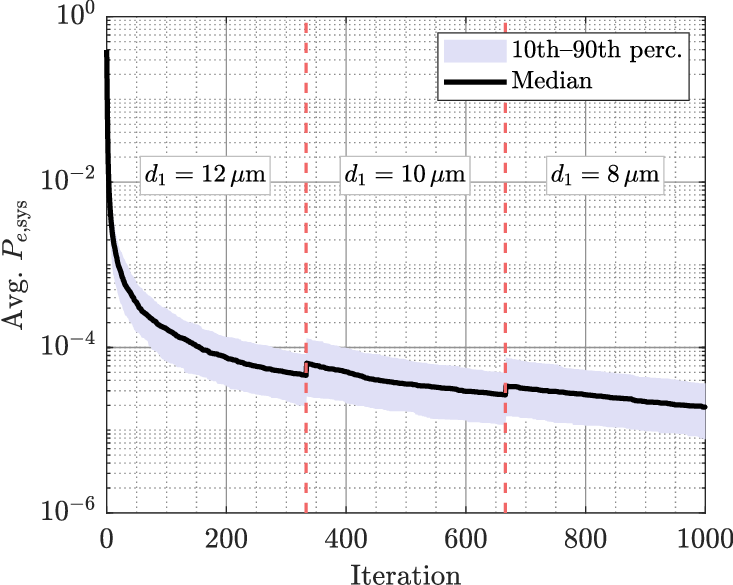}
        \caption{Distance $d_1$ varied from $\qty{12}{\micro\meter}$ to $\qty{8}{\micro\meter}$.}
        \label{fig:continuous:subfig:dist}
    \end{subfigure}
    \begin{subfigure}{\linewidth}
    \centering
        \includegraphics[width=0.8\linewidth]{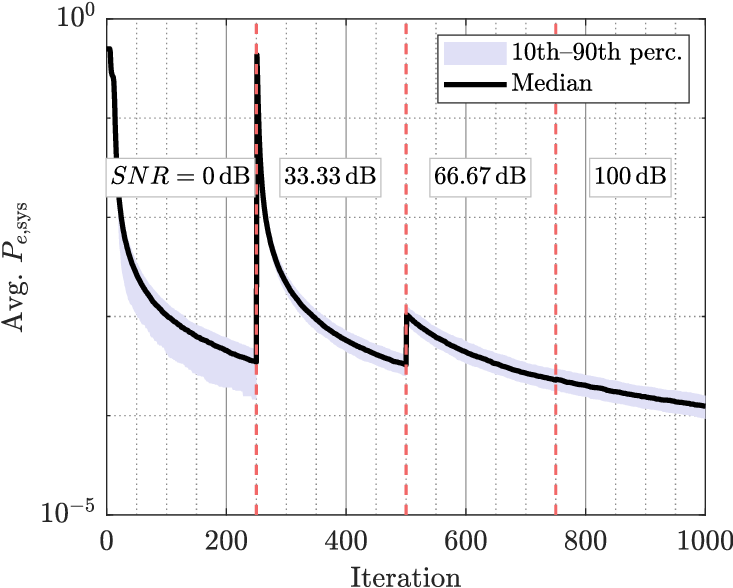}
        \caption{SNR varied from 0 to 100 dB.}
        \label{fig:continuous:subfig:snr}
    \end{subfigure}
    \caption{Average \ac{BEP}, $P_\mathrm{e,sys}$ (black line), and 10th-90th percentile range (shaded area), across 1000 iterations of the \ac{DBMC-aNOMAly} protocol. Results for variations in distance and SNR. The time of the change is indicated by the vertical red dashed lines. $K=2$. All other parameters according to the default in Table~\ref{tab:protocol_parameters}.}
    \label{fig:continuous}
\end{figure}

Lastly, we demonstrate the robustness of the \ac{DBMC-aNOMAly} protocol against fluctuations during the runtime, as would be observed in real moving and changing systems.
In Fig.~\ref{fig:continuous}, we consider two cases: Firstly, Fig.~\ref{fig:continuous:subfig:dist} varies the distance of TX$_1$ intermittently, as if it were moving. Secondly, Fig.~\ref{fig:continuous:subfig:snr} changes the \ac{SNR} at multiple points, emulating varying channel conditions and different sources of noise.

Starting with Fig.~\ref{fig:continuous:subfig:dist}, the distance $d_1$ of TX$_1$ from the \ac{RX} is adjusted from $\qty{12}{\micro\meter}$ to $\qty{8}{\micro\meter}$ in two steps at equidistant points across the 1000 iterations. The graph shows the median, as well as a 10th-90th percentile range to show the possible impact on the distribution.
The figure shows that the changes in distance have little impact on the $P_\mathrm{e,sys}$ trajectory as well as the percentile range, indicating the optimization procedure can deal with it without issues.

For Fig.~\ref{fig:continuous:subfig:snr}, the SNR was varied between 0 and 100 dB via three step-wise adjustments across 1000 iterations. In this case, we see an effect on the $P_\mathrm{e,sys}$ trajectory for the first two steps. For example, in the first step, the SNR changes from 0 to approximately 33 dB, which represents a significant shift in the received number of molecules due to the reduction in noise. Therefore, the current detection thresholds will be substantially mismatched. We can observe that the \ac{DBMC-aNOMAly} protocol is able to reduce the error to the pre-change levels before the next change occurs. In particular, the error drops quickly during the first few iterations after the initial change.
The last step from 66 to 100 dB represents only a small change in absolute noise level, to which the algorithm can adapt almost without visible disruption.
The percentile range shows that the abrupt change does not cause an increase in the variability of the optimization outcome, but rather that almost all trajectories follow a very similar path.

\section{Conclusion and Further Work}\label{sec:conclusion}

This paper presented an asynchronous \ac{NOMA}-based system model for \ac{DBMC} networks with $K$ \acp{TX} and one \ac{RX}, motivated by future \ac{IoBNT} networking scenarios. We derived the analytical \ac{BEP} and evaluated the effects of network size, noise level, and communication parameters. Detection thresholds, emitted molecule counts, and synchronization offsets were identified as the main performance factors. Compared with \ac{TDMA} and \ac{MDMA}, \ac{DBMC-NOMA} can match the upper-bound \ac{MDMA} performance if thresholds are optimized and worst-case offset configurations are avoided. We then proposed and evaluated DBMC-aNOMAly, a pilot-symbol-based optimization protocol for \ac{DBMC-NOMA}. Using three simple \ac{CRN}-compatible algorithms, the protocol optimizes key communication parameters while using a \ac{WCAM} to avoid worst-case offsets and reduce system \ac{BEP}. The results show robustness across channel conditions, network sizes, and sampling jitter. The end-to-end analysis further shows that these gains remain beneficial after accounting for pilot overhead, and the protocol can adapt to changing conditions.

We aim to address explicit modeling of the protocol as a chemical reaction network in future work, as well as the initial network setup and sampling procedure. Additionally, the \ac{DBMC-NOMA} scheme should be evaluated experimentally to validate analytical and simulation-based results.

% \section*{Acknowledgments}
% This should be a simple paragraph before the References to thank those individuals and institutions who have supported your work on this article.

% {\appendix[Proof of the Zonklar Equations]
% Use $\backslash${\tt{appendix}} if you have a single appendix:
% Do not use $\backslash${\tt{section}} anymore after $\backslash${\tt{appendix}}, only $\backslash${\tt{section*}}.
% If you have multiple appendixes use $\backslash${\tt{appendices}} then use $\backslash${\tt{section}} to start each appendix.
% You must declare a $\backslash${\tt{section}} before using any $\backslash${\tt{subsection}} or using $\backslash${\tt{label}} ($\backslash${\tt{appendices}} by itself
%  starts a section numbered zero.)}


\bibliographystyle{IEEEtran}
\bibliography{references}

% \newpage

 

% \begin{IEEEbiography}[{\includegraphics[width=1in,height=1.25in,clip,keepaspectratio]{fig1}}]{Alexander Wietfeld}
% \end{IEEEbiography}


% \end{IEEEbiography}

\vfill

\end{document}

%% file: abbreviations.tex
\newacronym{DBMC}{DBMC}{diffusion-based molecular communication}
\newacronym{TDMA}{TDMA}{time-division multiple access}
\newacronym{MDMA}{MDMA}{molecule-division multiple access}
\newacronym{NOMA}{NOMA}{non-orthogonal multiple access}
\newacronym{ADMA}{ADMA}{amplitude-division multiple access}
\newacronym{MA}{MA}{multiple access}
\newacronym{BEP}{BEP}{bit error probability}
\newacronym{TX}{TX}{transmitter}
\newacronym{RX}{RX}{receiver}
\newacronym{ISI}{ISI}{inter-symbol interference}
\newacronym{MAI}{MAI}{multiple-access interference}
\newacronym{SIC}{SIC}{successive interference cancellation}
\newacronym{OOK}{OOK}{on-off-keying}
\newacronym{MCS}{MCS}{Monte Carlo simulation}
\newacronym{SNR}{SNR}{signaling-molecule-to-noise ratio}
\newacronym{MC}{MC}{molecular communication}
\newacronym{IoBNT}{IoBNT}{Internet of Bio-Nano-Things}
\newacronym{BNM}{BNM}{Bio-Nano-Machine}
\newacronym{DBMC-NOMA}{DBMC-NOMA}{NOMA for DBMC networks}
\newacronym{UCA}{UCA}{uniform concentration assumption}
\newacronym{CRN}{CRN}{chemical reaction network}
\newacronym{SCRN}{SCRN}{stochastic chemical reaction network}
\newacronym{DBMC-aNOMAly}{DBMC-aNOMAly}{asynchronous NOMA with pilot-symbol optimization protocol for DBMC}
\newacronym{EM}{EM}{electromagnetic}
\newacronym{IoT}{IoT}{Internet of Things}
\newacronym{MI}{MI}{mutual information}
\newacronym{WCAM}{WCAM}{worst-case-offset avoidance mechanism}